\documentclass[11pt,a4paper]{article}
\usepackage{jheppub}
\usepackage{feynarts}
\usepackage{graphics}
\usepackage{amsmath}
\usepackage{lscape}
\usepackage{wasysym}
\usepackage{rotating}
\usepackage{url}

\newcommand{\GeV}{\mathrm{GeV}}

\newcommand{\pba}{\mathrm{pb}}

\def\mathswitch#1{\relax\ifmmode#1\else$#1$\fi}
\def\mathswitchr#1{\relax\ifmmode{\mathrm{#1}}\else$\mathrm{#1}$\fi}
\def\mathswitchit#1{\relax\ifmmode{#1}\else$#1$\fi}
\newcommand{\muF}{\mu_{\mathrm{F}}}

\newcommand{\PW}{\mathswitchr W}
\newcommand{\PZ}{\mathswitchr Z}

\newcommand{\PH}{\mathswitchr H}

\newcommand{\Pe}{\mathswitchr e}

\newcommand{\pt}{p_{\mathrm{T}}}

\renewcommand{\O}{{\cal O}}

\newcommand{\ri}{{\mathrm{i}}}

\newcommand{\rT}{{\mathrm{T}}}

\newcommand{\rL}{{\mathrm{L}}}

\newcommand{\rd}{{\mathrm{d}}}

\newcommand{\M}{{\cal {M}}}

\newcommand{\cut}{{\mathrm{cut}}}

\newcommand{\EW}{{\mathrm{EW}}}

\newcommand{\LO}{{\mathrm{LO}}}
\newcommand{\NLO}{{\mathrm{NLO}}}

\def\Re{\mathop{\mathrm{Re}}\nolimits}

\def\draftdate{\relax}
\def\mda{\relax}
\def\mua{\relax}
\def\mla{\relax}
\def\draft{
\def\thtystars{******************************}
\def\sixtystars{\thtystars\thtystars}
\typeout{}
\typeout{\sixtystars**}
\typeout{* Draft mode!
         For final version remove \protect\draft\space in source file *}
\typeout{\sixtystars**}
\typeout{}
\def\draftdate{\today}
\def\mua{\marginpar[\boldmath\hfil$\uparrow$]%
                   {\boldmath$\uparrow$\hfil}%
                    \typeout{marginpar: $\uparrow$}\ignorespaces}
\def\mda{\marginpar[\boldmath\hfil$\downarrow$]%
                   {\boldmath$\downarrow$\hfil}%
                    \typeout{marginpar: $\downarrow$}\ignorespaces}
\def\mla{\marginpar[\boldmath\hfil$\rightarrow$]%
                   {\boldmath$\leftarrow $\hfil}%
                    \typeout{marginpar: $\leftrightarrow$}\ignorespaces}
\def\Mua{\marginpar[\boldmath\hfil$\Uparrow$]%
                   {\boldmath$\Uparrow$\hfil}%
                    \typeout{marginpar: $\Uparrow$}\ignorespaces}
\def\Mda{\marginpar[\boldmath\hfil$\Downarrow$]%
                   {\boldmath$\Downarrow$\hfil}%
                    \typeout{marginpar: $\Downarrow$}\ignorespaces}
\def\Mla{\marginpar[\boldmath\hfil$\Rightarrow$]%
                   {\boldmath$\Leftarrow $\hfil}%
                    \typeout{marginpar: $\Leftrightarrow$}\ignorespaces}
\overfullrule 5pt
\oddsidemargin -15mm
\marginparwidth 29mm
}

\makeatother

\title{Vector-boson pair production at the LHC to
  $\boldsymbol{\mathcal{O}(\alpha^3)}$ accuracy}

\author{Anastasiya Bierweiler,}
\author{Tobias Kasprzik,}
\author{Johann H.~K\"uhn}

\affiliation{ 
Karlsruhe Institute of Technology (KIT), 
Institut f\"ur Theoretische Teilchenphysik,\\
D-76128 Karlsruhe, Germany}

\emailAdd{kasprzik@particle.uni-karlsruhe.de}
\date{\today}

\abstract{Building on earlier work on electroweak corrections to W-pair
  production, the first calculation of the full electroweak one-loop
  corrections to on-shell ZZ, W$^{\pm}$Z and $\gamma\gamma$ production
  at hadron colliders is presented, explicitly taking into account the
  full vector-boson mass dependence. As a consequence, our results are
  valid in the whole energy range probed by LHC experiments. Until now,
  the electroweak corrections have only been known in dedicated
  high-energy approximations limited to a specific kinematic regime, in
  particular requiring high boson transverse momenta. Therefore, our
  results comprise an important and so far missing ingredient to improve
  on the theory predictions for these fundamental Standard-Model
  benchmark processes also at intermediate energies and small scattering
  angles, where actually the bulk of events is located. In case of
    Z-pair production we have also included the leptonic decays and the
    associated weak corrections in our analysis. For this particular
    channel, corrections of about $-4\%$ are observed even close to the
  production threshold. For hard scattering processes with momentum
  transfers of several hundred GeV one finds large negative corrections
  which may amount to several tens of percent and lead to significant
  distortions of transverse-momentum and rapidity distributions.

\vspace{1cm}
\begin{flushright}
\emph{SFB/CPP-13-35\\
TTP13-019\\
LPN13-031}
\end{flushright}}

\begin{document}
\maketitle 
\flushbottom

\section{Introduction}
A profound understanding of vector-boson pair production processes at
the LHC is desirable for various reasons. Such processes not only
contribute an important irreducible background to Standard-Model (SM)
Higgs production at moderate energies, but will also provide deeper
insight into the physics of the weak interaction at the high-energy
frontier, possibly even allowing for the discovery of BSM
physics. Consequently, great effort has been made during the last
years to push the theory predictions for this process class to a new
level, where, besides the dominating QCD corrections, also electroweak
(EW) effects have been studied extensively (see, e.g.,
ref.~\cite{Bierweiler:2012kw} and references therein).

Extending our work on EW effects in W-pair production at the LHC~,
we present\footnote{Preliminary results of this investigation have been
  presented in ref.~\cite{Bierweiler:2012qq}.} corresponding results for
on-shell W$^\pm$Z, ZZ and $\gamma\gamma$ pair production in the SM. We
will restrict ourselves to pair production through quark--antiquark
annihilation. Photon--photon collisions do not contribute to WZ
production (in contrast to the case of W pairs), are of higher order for
Z-pair and $\gamma$-pair production and will not be discussed
further. Also gluon fusion is, evidently, irrelevant for WZ
production. For the case of W-pair production we have demonstrated that
gluon fusion amounts to order of 5\% relative to the quark--antiquark
annihilation with decreasing importance for increasing transverse
momenta. Gluon fusion, furthermore, does not lead to a strong
modification of the angular and rapidity distributions of the W
bosons. A qualitatively similar behaviour is expected for ZZ and
$\gamma\gamma$ production through gluon fusion, which therefore will not
be discussed further. Also QCD corrections for WZ and ZZ/$\gamma\gamma$
production are expected to be similar to those for W pairs discussed in
ref.~\cite{Bierweiler:2012kw} and will not be analyzed in the present
paper, which, instead, will be entirely devoted to genuine EW
corrections. 

Earlier papers have emphasized the drastic influence of the EW
corrections on cross sections and distributions in the high-$p_{\rT}$
region. However, only pair-production channels with at least one massive
gauge boson were considered~\cite{Accomando:2001fn, Accomando:2005ra,
  Accomando:2004de} while a phenomenological analysis of the
$\gamma\gamma$ channel is still missing. To close this gap, the first
computation of the full EW corrections to photon-pair hadroproduction is
presented in this work.  Moreover, the above papers neglect terms of
order $M_{V}^2/\hat{s}$. Also this issue will be addressed in the
present paper. All the processes under consideration are affected by
large corrections, amounting to several tens, sometimes up to fifty
percent. (This fact has even triggered studies of
next-to-next-to-leading logarithmic corrections for W-pair production at
the LHC~\cite{Kuhn:2011mh}, which, however, where not extended to Z-pair
or WZ production.) Nevertheless, significant differences are observed
between the different final states, as far as the size of the
corrections is concerned. We, furthermore, investigate rapidity
distributions, both for small and for large invariant masses of the
diboson system and observe significant distortions that could be
misinterpreted as anomalous triple-boson couplings.  The full mass
dependence, namely terms proportional to powers of $M_V^2/\hat{s},$ is
consistently accounted for to obtain results valid in the whole energy
range probed by LHC experiments, in particular for $V$-pairs produced
near threshold\footnote{Although EW corrections are expected to be small
  at low energies, for Z-pair production a shift of about $-4\%$ is
  observed even close to threshold, whereas the corresponding
  corrections are below 1 percent in the remaining cases.} or at low
transverse momenta. Therefore, our results are complementary to those
presented in refs.~\cite{Accomando:2001fn, Accomando:2004de}, where only
logarithmic corrections were considered, but the leptonic decays of the
vector bosons and related off-shell effects were included in a
double-pole approximation. Comparing both approaches, we try to estimate
the remaining theoretical uncertainties related to EW corrections in
this important process class.

To investigate this aspect in more detail, also the combined production
and decay process for Z pairs is calculated, including NLO
corrections. To simplify the discussion, only the four-lepton state
$\Pe^+\Pe^-\mu^+\mu^-$ is considered. Excellent agreement with the
results of Accomando et al.\ is observed in the region where the
high-energy approximation employed in ref.~\cite{Accomando:2004de} is
valid. We, furthermore, investigate the contributions from the different
helicity configurations including NLO corrections and observe that the
transverse polarization not only dominates completely at large
$p_{\rT}$, it is also the most prominent configuration as far as the
total cross section is concerned.
 
In addition, we also study the impact of
real gauge-boson radiation leading to three-boson final states. In
principle, these configurations might compensate some of the negative
contributions from virtual corrections, in practice, however, real
radiation is significantly smaller than virtual contributions.

\section{Details of the calculation}

\subsection{Contributions at leading order}
At the LHC, WW, W$^\pm$Z, ZZ and $\gamma\gamma$ production
is, at lowest order $\mathcal{O}(\alpha^2)$, induced by the partonic processes
\begin{subequations}
\label{eq:LO}
\begin{eqnarray}
  q\bar{q} &\to& \mathrm{W^-W^+} \quad (q = \mathrm{u,d,s,c,b})\,, \\
  \mathrm{u}_i\mathrm{\bar{d}}_j &\to& \mathrm{W^+Z}\,, \quad \mathrm{\bar{u}}_i\mathrm{d}_j \to \mathrm{W^-Z} \quad (i = 1,2;\;
  j=1,2,3)\,, \\
  q\bar{q} &\to& \mathrm{ZZ}\,,\\
  q\bar{q} &\to& \mathrm{\gamma\gamma}\,.
\end{eqnarray}
\end{subequations}
where the corresponding LO partonic cross
sections are evaluated according to 
\begin{equation}
\hat{\sigma}^{q\bar{q}' \to V_1V_2}_{\LO} =
\frac{1}{2\hat{s}N_{q\bar{q}'}}\int\rd\Phi(V_1V_2)\sum_{\mathrm{col}}\sum_{\mathrm{spin}}\sum_{\mathrm{pol}}
|\M^{q\bar{q}' \to V_1V_2}_0|^2\,,
\end{equation}
with the tree-level helicity amplitudes $\M^{q\bar{q}' \to V_1V_2}_0$,
the two-particle phase-space measure $\rd\Phi(V_1V_2)$ and the averaging
factor $N_{q\bar{q}'} = 36$.  It is understood that for ZZ and
  $\gamma\gamma$ production the cross sections receive an additional
  symmetry factor of $1/2$. Note that one finds $\int \rd \xi_V \,(\rd
  \sigma^{\mathrm{pp} \to VV}/\rd \xi_V) = 2 \sigma^{\mathrm{pp} \to VV}$
  for $\xi = p_{\rT}$ or  $\xi = y$ for these particular channels.

\subsection{Electroweak radiative corrections}
To allow for consistent predictions with full $\mathcal{O}(\alpha^3)$
accuracy, virtual EW corrections  as well as real corrections due to
photon radiation have to be considered. The evaluation of the radiative
corrections is based on the well-established {\tt
  Feyn\-Arts/FormCalc/LoopTools} setup~\cite{Kublbeck:1990xc,
  Hahn:2000kx, Hahn:1998yk, Hahn:2001rv, vanOldenborgh:1989wn}, and all
processes have been independently cross-checked by a setup based on {\tt
  QGraf}~\cite{Nogueira:1991ex} and {\tt Form}~\cite{Vermaseren:2000nd}.

To considerably reduce the computational effort, light quark masses are
neglected whenever possible. However, soft and collinear singularities
occurring in intermediate steps of the calculation are regularized by
small quark masses $m_q$ and an infinitesimal photon mass $\lambda$,
generating unphysical $\ln m_q$ and $\ln \lambda$ terms. To allow for a
numerically stable evaluation of those infrared(IR)-divergent parts of the
cross sections related to real radiation, the phase-space slicing method
is adopted as detailed in ref.~\cite{Dittmaier:2001ay} for mass regularization. Finally,
adding real and virtual contributions, the regulator-mass dependence
drops out in any properly defined physical result. However, in complete
analogy to QCD, residual collinear singularities attributed to
initial-state (IS) radiation survive and have to be absorbed in
renormalized PDFs in a proper factorization procedure adding the
collinear counterterm defined by eq.\ (3.16) of ref.~\cite{Diener:2003ss}. In the present
computation, we apply the $\overline{\mathrm{MS}}$ factorization scheme
for the QED factorization.

The input parameters to be specified in Section~\ref{se:setup} are
renormalized in a modified on-shell scheme~\cite{Denner:1991kt}, where
the Fermi constant $G_\mu$ is used instead of $\alpha(0)$ to effectively
account for universal corrections induced by the running of
$\alpha(\mu)$ to the weak scale~\cite{Dittmaier:2001ay}. However, for
the computation of the $\gamma\gamma$ channel, $\alpha(0)$ is used as
input for the EW coupling, since the corresponding radiative corrections
do not receive universal contributions related to the running of the
 coupling constant. According to the previous considerations, the
partonic cross section at $\O(\alpha^3)$ accuracy may be written as
\begin{equation}
\hat{\sigma}^{q\bar{q}' \to V_1V_2(\gamma)}_{\NLO} = \hat{\sigma}^{q\bar{q}' \to
  V_1V_2}_{\LO} + \hat{\sigma}^{q\bar{q}' \to
  V_1V_2\gamma}_{\LO} +  \hat{\sigma}^{q\bar{q}' \to
  V_1V_2}_{\mathrm{virt}}\,, 
\end{equation}
where the different NLO contributions are given by
\begin{eqnarray}
\hat{\sigma}^{q\bar{q}' \to V_1V_2\gamma}_{\LO} &=&
\frac{1}{2\hat{s}N_{q\bar{q}'}}\int\rd\Phi(V_1V_2\gamma)\sum_{\mathrm{col}}\sum_{\mathrm{spin}}\sum_{\mathrm{pol}}
|\M^{q\bar{q}' \to V_1V_2\gamma}_0|^2\,,\label{eq:real_gamma}\\
\hat{\sigma}^{q\bar{q}' \to V_1V_2}_{\mathrm{virt}} &=&
\frac{1}{2\hat{s}N_{q\bar{q}'}}\int\rd\Phi(V_1V_2)\sum_{\mathrm{col}}\sum_{\mathrm{spin}}\sum_{\mathrm{pol}}
2 \Re \left\{ (\M^{q\bar{q}' \to V_1V_2}_0)^* \,\M^{q\bar{q}' \to V_1V_2}_1\right\}\,,\label{eq:virt}
\end{eqnarray}
with the properly renormalized one-loop amplitudes $\M^{q\bar{q}' \to
  V_1V_2}_1$. The hadronic results at the LHC are then obtained by convoluting the partonic
cross sections with appropriately chosen PDFs and summing incoherently
over all contributing channels, 
\begin{equation}
\sigma^{\mathrm{pp} \to V_1V_2(\gamma)}_{\NLO} = \int_{\tau_0}^1 \rd \tau \int_\tau^1 \frac{\rd x_b}{x_b}
  \, \sum_{q,\bar{q}'} 
  f_{q/\mathrm{p}}(x_a,\muF^2)f_{\bar{q}'/\mathrm{p}}(x_b,\muF^2)\,\hat{\sigma}_{\NLO}^{q\bar{q}' \to
    V_1V_2(\gamma)}(\tau s,\muF^2)\,,
\end{equation}
where the hadronic CM energy $s$ is related to $\hat{s}$ via $\hat{s} =
\tau s$, with $\tau = x_ax_b$.  The kinematic production threshold of a
vector-boson pair in the final state is reflected by the choice of the
lower integration boundary $\tau_0 = (M_{V_1}+M_{V_2})^2/s$,
corresponding to a minimal partonic CM energy of $\hat{s}_0 = \tau_0
s$. The factorization scale $\muF$ enters the partonic cross section
through the redefinition of the PDFs in the QED factorization procedure
described above.

\subsection{Radiation of massive gauge bosons}
\label{se:real_rad_def}
As in ref.~\cite{Bierweiler:2012kw} we study the effect of massive boson
radiation on the massive vector-boson pair production cross section,
where the additional massive boson is treated fully exclusively in the
event selection to allow for a robust estimate of the corresponding
phenomenological effects. This is motivated by the possibility that the
logarithmically enhanced positive contributions from the radiation of an
additional soft or collinear massive gauge boson might compensate the
negative virtual corrections, as has been argued recently for
  Z+jet production~\cite{Stirling:2012ak}.

 To be specific, we take into account
the following partonic channels, where the corresponding partonic cross
sections have to be computed according to
\begin{equation}
  \hat{\sigma}_{\LO}^{q\bar{q}' \to V_1V_2V_3} = \frac{1}{2\hat{s}N_{q\bar{q}'}}\int\rd\Phi(V_1V_2V_3)\sum_{\mathrm{col}}\sum_{\mathrm{spin}}\sum_{\mathrm{pol}}
  |\M^{q\bar{q}' \to V_1V_2V_3}_0|^2\,.
\end{equation}
In case of ZZ pair production, the contributions from the processes
\begin{eqnarray}
q\bar{q} &\to& \PZ\PZ\PZ\,,\nonumber \\
q\bar{q}'&\to& \PZ\PZ\PW^{\pm}
\end{eqnarray} 
have to be considered, where the two Zs with highest $\pt$ have to
fulfill the LO cuts to be specified in section~\ref{se:setup}, while the
third massive particle is treated inclusively.  To assess the
real-radiation effects in $\PW^{\pm}\PZ$ production, we compute the
cross sections for
\begin{eqnarray}
q\bar{q} &\to& \PW^{\pm}Z\PW^{\mp}\,,\nonumber \\
q\bar{q}'&\to& \PW^{\pm}\PZ\PZ\,.
\end{eqnarray} 
In the first case the W$^\mp$ is treated inclusively, in the
second case the Z with lowest $\pt$.  In section~\ref{se:real_rad}, numerical results will be
presented  for
massive boson radiation normalized to the LO $q\bar{q}$-induced
  pair-production channels, 
\begin{equation}
\delta_{V_1V_2V_3} = \sigma^{{\mathrm{pp} \to V_1V_2V_3}}_{\LO} /
\sigma^{\mathrm{pp} \to V_1V_2}_{\LO} - 1\,, 
\end{equation}
as well as for hard photon radiation (i.e.\ $p_{\mathrm{T},\gamma} > 15$
GeV, $|y_{\gamma}| < 2.5$), 
\begin{equation}
\delta_{V_1V_2\gamma} = \sigma^{{\mathrm{pp} \to V_1V_2\gamma}}_{\LO} /
\sigma^{\mathrm{pp} \to V_1V_2}_{\LO} - 1\,.  
\end{equation}

\section{Numerical results}
In this section, for the first time the full EW corrections to Z-
and $\gamma$-pair production as well as $\PW^\pm \PZ$ production are
presented. Most of the discussion is focussed on LHC8 and LHC14.
For completeness, we also show cross sections for these processes at the
Tevatron with a CM energy of $\sqrt{s} = 1.96$ TeV. The corresponding
results for W-pair production originally presented in
ref.~\cite{Bierweiler:2012kw} are given for comparison. In the
following, the relative effects $\delta_{\EW}$ of the EW corrections are
defined as
\begin{equation}
\label{eq:delew}
\delta_{\EW} =
\frac{\sigma^{\mathrm{pp} \to V_1V_2(\gamma)}_{\NLO}}{\sigma^{\mathrm{pp} \to V_1V_2}_{\LO}} - 1\,. 
\end{equation}

\subsection{Input and setup}
\label{se:setup}
For the computation presented here the same setup as specified in
ref.~\cite{Bierweiler:2012kw} is applied. To be specific, we use the
following SM input parameters for the numerical analysis,
\begin{equation}\arraycolsep 2pt
\begin{array}[b]{lcllcllcl}
G_{\mu} & = & 1.16637 \times 10^{-5} \;\mathrm{GeV}^{-2}, \quad & & & & &  \\
M_{\mathrm{W}} & = & 80.398\;\mathrm{GeV}, & M_{\mathrm{Z}} &
= & 91.1876\;\mathrm{GeV}, \quad & & & \\ 
 M_\PH & = & 125\;\mathrm{GeV}, & M_{{\rm t}} & = & 173.4\;\mathrm{GeV}\,. & & & 
\end{array}
\label{eq:SMpar}
\end{equation}
For the evaluation of all tree-level contributions we assume a
block-diagonal CKM matrix with
\begin{equation}
\label{eq:CKM}
|V_{\mathrm{ud}}| = |V_{\mathrm{cs}}| = 0.974\,,\quad |V_{\mathrm{us}}|
= |V_{\mathrm{cd}}| = \sqrt{1 - |V_{\mathrm{ud}}|^2}\,.
\end{equation}
Ignoring, furthermore, quark masses within the first two families, both
tree-level and one-loop predictions for ZZ and $\gamma\gamma$ are
equivalent to those without quark mixing. As a consequence of the
smallness of the bottom-quark PDF the tree-level contribution from
$\mathrm{b\bar{b}}$ annihilation to ZZ or $\gamma\gamma$ is small to
start with. In addition, the non-diagonal CKM elements involving b
quarks are small, and the ansatz~\eqref{eq:CKM} is well justified.  As a
consequence, $\mathrm{b\bar{b}} \to \mathrm{ZZ}$ or $\gamma\gamma$ can
safely be handled within the third family.\footnote{We point out
    that a non-vanishing top-quark mass is consistently included in the
    computation of the one-loop contributions discussed in this paper.}
The situation is different for the WZ channel. In this case, the
interplay between CKM angles and PDFs leads to a shift of the tree-level
prediction of about one percent. For the radiative corrections the CKM
matrix can, therefore, still be set to unity.

In the on-shell scheme applied in our computation, the weak mixing angle
$\cos^2\theta_{\mathrm{w}} = M_{\mathrm{W}}^2/M_{\mathrm{Z}}^2$ is a
derived quantity.  For the computation of the processes~\eqref{eq:LO}
and the corresponding EW radiative corrections, we use the
MSTW\-2008\-LO PDF set~\cite{Martin:2009iq} in the LHAPDF
setup~\cite{Whalley:2005nh}. In order to consistently include
$\mathcal{O}(\alpha)$ corrections, in particular real radiation with the
resulting collinear singularities, PDFs in principle should take these
QED effects into account. Such a PDF analysis has been performed in
ref.~\cite{Martin:2004dh}, and the $\mathcal{O}(\alpha)$ effects are
known to be small, as far as their effect on the quark distribution is
concerned~\cite{Roth:2004ti}. In addition, the currently available PDFs
incorporating $\mathcal{O}(\alpha)$ corrections~\cite{Martin:2004dh}
include QCD effects at NLO, whereas our EW analysis is LO with respect
to perturbative QCD only. For these reasons, the MSTW2008LO set is used
as our default choice for the quark-induced processes. Our default
choice for the factorization scale is the average of the vector-boson
transverse masses
\begin{equation}
\mu_{{\rm F}} = \overline{m_{\rT}} =
\frac{1}{2}\left(\sqrt{M_{V_1}^2+p_{\rT,V_1}^2}+\sqrt{M_{V_2}^2+p_{\rT,V_2}^2}\right)\,.
\end{equation}
A similar scale choice was taken in ref.~\cite{Accomando:2004de} for the
computation of the EW corrections to four-lepton production at the
LHC. Yet we point out that the relative EW corrections, which are the
main subject of this paper, only depend on the choice of
  $\mu_{\mathrm{F}}$ at the subpercent level even for large transverse
  momenta.

 In our default setup, we require a minimum transverse
momentum and a maximum rapidity for the final-state vector bosons,
\begin{equation}\label{eq:defcuts}
 p_{\rT,V_i} > 15\;\mathrm{GeV}\,,\quad |y_{V_i}|<2.5\,, \quad i=1,2\,,
\end{equation}
to define a $V$-boson pair production event.  Thereby we exclude events
where the bosons are emitted collinear to the initial-state partons,
which for the $\gamma\gamma$ channel would inevitably lead to collinear
singularities in the LO cross section. However, for final states with
massive gauge bosons selected numerical results will also be presented
without applying the above cuts.

For the definition of a two-photon final state we require at least two
visible photons fulfilling the acceptance cuts~\eqref{eq:defcuts}. If
additional photon bremsstrahlung is present, any further phase-space
cuts will only be applied to the two visible photons with highest
$p_{\rT}$, while the third $\gamma$ is treated inclusively to ensure IR
safety.

\subsection{Leading-order cross sections}
\begin{figure}
\begin{center}
\includegraphics[width = 1.0\textwidth]{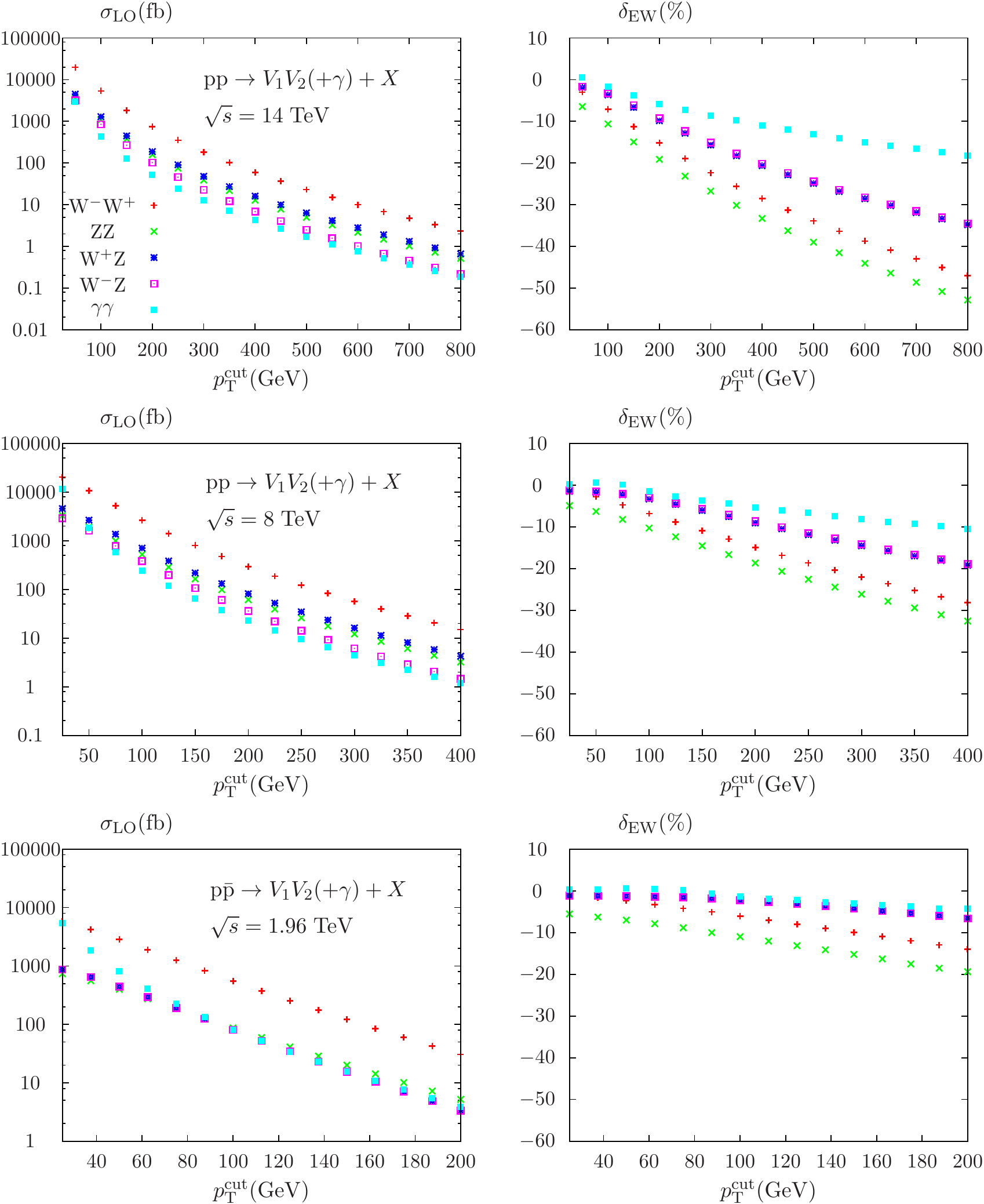}
\end{center}
\caption{\label{fi:ptcut_vv}Integrated LO cross sections (left) and relative EW corrections
  (right) evaluated with our default setup for different cuts on the
  transverse momenta of the final-state bosons.}
\end{figure}

\begin{figure}
\begin{center}
\includegraphics[width = 1.0\textwidth]{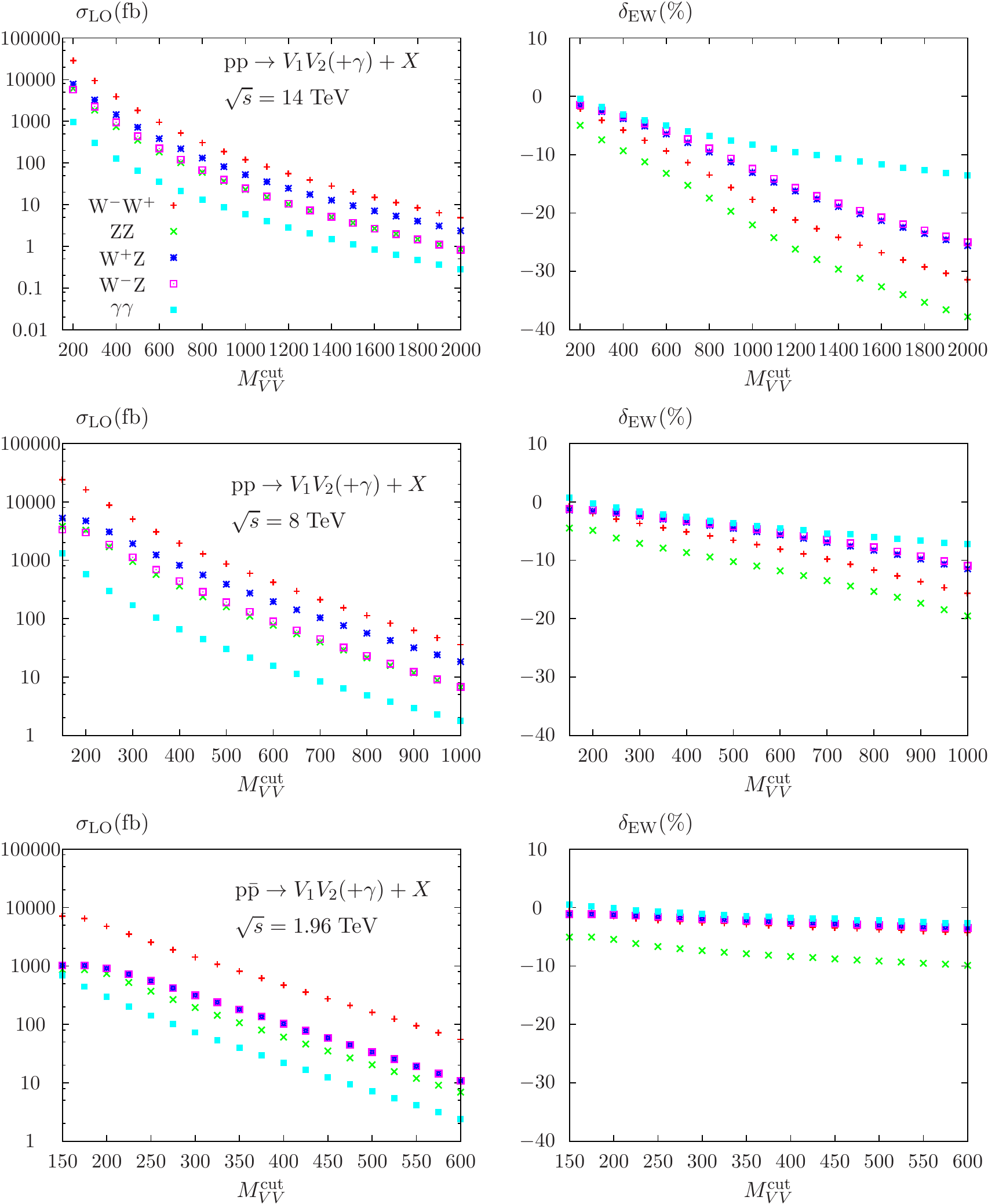}
\end{center}
\caption{\label{fi:mcut_vv}Integrated LO cross sections (left) and relative EW corrections
  (right) evaluated with our default setup for different cuts on the
  invariant mass of the final-state bosons. The respective results are
  presented for LHC14 (top), LHC8 (center) and the Tevatron (bottom).}
\end{figure}

\begin{table}
\footnotesize
\begin{equation}
\begin{array}{|c|| c|c|c|c|c|c|c|c|c|c|}
\hline
 & \multicolumn{2}{|c}{\PW^-\PW^+} &  \multicolumn{2}{|c}{\PZ\PZ} &  \multicolumn{2}{|c}{\PW^+\PZ} &  \multicolumn{2}{|c}{\PW^-\PZ}& \multicolumn{2}{|c|}{\gamma\gamma}  \nonumber \\ 
\mbox{default cuts} & \sigma_\LO\,(\pba) & \delta_{\EW}(\%) &  \sigma_\LO\,(\pba) & \delta_{\EW}(\%) &  \sigma_\LO\,(\pba) &
 \delta_\EW(\%) &  \sigma_\LO\,(\pba) & \delta_\EW(\%) &  \sigma_\LO\,(\pba) & \delta_\EW (\%) \nonumber\\
\hline\hline
\mathrm{LHC8}    & 23.99 & -0.7 & 3.810  & - 4.4 & 5.256 & -1.4 &  3.343 & -1.2 &41.38 & 0.2 \nonumber\\
\mathrm{LHC14}   & 42.39 & -0.9 & 7.066  & - 4.5 & 8.677 & -1.5 &  6.463& -1.3 & 62.69&0.2 \nonumber\\
\hline
\mathrm{Tevatron}& 7.054 & -0.5 & 0.8624 & - 4.9 & 1.023 & -1.1 &  1.023 & -1.1 & 19.21 &0.2 \nonumber\\
\hline
\end{array}
\end{equation}
\caption{\label{ta:totcs_def} Integrated leading-order cross sections
  and relative  EW corrections for the LHC and the Tevatron evaluated with the
  default setup defined in Section~\ref{se:setup}.}
\end{table}

\begin{table}
\begin{equation}
\begin{array}{|c||c|c|c|c|c|c|c|c|}
\hline
 &\multicolumn{2}{|c}{\PW^-\PW^+} & \multicolumn{2}{|c}{\PZ\PZ}&  \multicolumn{2}{|c}{\PW^+\PZ} & \multicolumn{2}{|c|}{\PW^-\PZ} \nonumber \\
\mbox{no cuts} & \sigma_\LO\,(\pba) & \delta_{\EW}(\%) &  \sigma_\LO\,(\pba) & \delta_{\EW}(\%) &  \sigma_\LO\,(\pba) &
\delta_\EW(\%) &  \sigma_\LO\,(\pba) & \delta_\EW(\%)  \nonumber\\
\hline\hline
\mathrm{LHC8}    & 35.51  & -0.4  & 5.064  & - 4.1 & 8.273 & -1.4 & 4.643 & -1.3  \nonumber\\
\mathrm{LHC14}   & 75.02  & -0.4  & 11.02  & - 4.2 & 17.11 & -1.4 & 10.65 & -1.3  \nonumber\\
\hline
\mathrm{Tevatron}& 7.916  & -0.2  & 0.9466 & - 4.7 & 1.123 & -1.1 & 1.123 & -1.1  \nonumber\\
\hline
\end{array}
\end{equation}
\caption{\label{ta:totcs_nocut} Total leading-order cross sections
  and relative  EW corrections for the LHC and the Tevatron evaluated
  without any phase-space cuts.}
\end{table}
\begin{table}
\footnotesize
\begin{equation}
\begin{array}{|c||c|c|c|c|c|c|c|c|}
\hline
 \multicolumn{9}{|c|}{\mathrm{pp} \to V_1V_2(+\gamma) + X \mbox{ at }
   \sqrt{s} = 14 \mbox{ TeV}} \\
\hline
 \mbox{default cuts} & \multicolumn{2}{|c}{\PZ\PZ} &  \multicolumn{2}{|c}{\PW^+\PZ} &  \multicolumn{2}{|c}{\PW^-\PZ}& \multicolumn{2}{|c|}{\gamma\gamma}  \nonumber \\ 
p_{\rT}^{\cut}\,(\GeV) &  \sigma_\LO\,(\pba) & \delta_{\EW}(\%) &  \sigma_\LO\,(\pba) &
\delta_\EW(\%) &  \sigma_\LO\,(\pba) & \delta_\EW(\%) &  \sigma_\LO\,(\pba) & \delta_\EW (\%) \nonumber\\
\hline\hline
50 & 3.660 & -6.3 & 4.498 & -1.9  & 3.228 & -1.7 & 2.979 & \phantom{-}0.6 \nonumber \\
100 & 1.087 & -10.4 & 1.296 & -3.7 & 0.849 & -3.3 & 0.432 & -1.6 \nonumber \\
250 & 7.495\times 10^{-2} & -23.0 & 90.56\times 10^{-2} & -12.9 &
4.583\times 10^{-2} & -12.3 & 2.451\times 10^{-2} & -7.0 \nonumber \\
500 & 49.89\times 10^{-4} & -38.9 & 63.64\times 10^{-4} & -24.9 &
24.79\times 10^{-4} & -24.4 & 16.95\times 10^{-4} & -13.0 \nonumber \\
750  & 72.16\times 10^{-5} & -50.6 & 92.43\times 10^{-5} & -33.3 &
31.25\times 10^{-5} & -33.0 & 25.66\times 10^{-5} & -17.3 \nonumber\\
1000 & 14.60\times 10^{-5} & -60.1 & 18.35\times 10^{-5} & -39.8 &
57.32\times 10^{-6} & -39.6 & 54.91\times 10^{-6} & -20.6 \nonumber \\
1250 & 35.28\times10^{-6}  & -68.4 & 42.92\times 10^{-6} & -45.1 &
12.86\times 10^{-6} & -45.1 & 14.15\times 10^{-6} & -23.5 \nonumber\\
1500 & 94.73\times 10^{-7} & -75.7 & 10.99\times 10^{-6} & -49.6 &
32.48\times 10^{-7} & -49.6 & 40.71\times 10^{-7} & -25.9  \nonumber\\
\hline
\end{array}
\end{equation}
\caption{\label{ta:totcs_ptcut} Integrated leading-order cross sections
  and relative  EW corrections at LHC14 for different cuts on
  the minimal boson transverse momenta.}
\end{table}

\begin{table}
\footnotesize
\begin{equation}
\begin{array}{|c||c|c|c|c|c|c|c|c|}
\hline
 \multicolumn{9}{|c|}{\mathrm{pp} \to V_1V_2(+\gamma) + X \mbox{ at }
   \sqrt{s} = 14 \mbox{ TeV}} \\
\hline
\mbox{default cuts} & \multicolumn{2}{|c}{\PZ\PZ} &  \multicolumn{2}{|c}{\PW^+\PZ} &  \multicolumn{2}{|c}{\PW^-\PZ}& \multicolumn{2}{|c|}{\gamma\gamma}  \nonumber \\ 
M_{VV}^{\cut}\,(\GeV) &  \sigma_\LO\,(\pba) & \delta_{\EW}(\%) &  \sigma_\LO\,(\pba) &
\delta_\EW(\%) &  \sigma_\LO\,(\pba) & \delta_\EW(\%) &  \sigma_\LO\,(\pba) & \delta_\EW (\%) \nonumber\\
\hline\hline
200 & 6.094  & -5.0  & 7.820 & -1.6 & 5.790 & -1.5 & 0.966 & -0.4 \nonumber\\
300 & 1.859  & -7.4  & 3.252 & -2.6 & 2.273 & -2.4 & 0.304 & -1.8 \nonumber\\
500 & 0.352  & -11.2 & 0.716 & -5.1 & 0.440 & -4.7 & 64.54\times
10^{-3} & -3.8 \nonumber\\
1000& 23.10\times 10^{-3} & -22.0 & 52.70\times 10^{-3} & -13.1 &
24.47\times 10^{-3} & -12.4 & 58.63\times 10^{-4} & -8.1 \nonumber \\
1500& 36.65\times 10^{-4} & -31.2 & 94.50\times 10^{-4} & -20.1 &
36.67\times 10^{-4} & -19.6 & 11.09\times 10^{-4} & -11.0 \nonumber \\
2000& 84.75\times 10^{-5} & -37.8 & 23.73\times 10^{-4} & -25.6 &
82.87\times 10^{-5} & -25.0 & 28.25\times 10^{-5} & -13.4 \nonumber \\
2500& 23.17\times 10^{-5} & -43.4 & 68.36\times 10^{-5} & -30.2 &
22.52\times 10^{-5} & -29.6 & 84.20\times 10^{-6} & -15.4 \nonumber \\
3000& 69.60\times 10^{-6} & -48.2 & 16.72\times 10^{-5} & -35.0 &
67.72\times 10^{-6} & -33.5 & 27.49\times 10^{-6} & -17.0 \nonumber \\
\hline
\end{array}
\end{equation}
\caption{\label{ta:totcs_mcut} Integrated leading-order cross sections
  and relative  EW corrections at LHC14 for different cuts on the
  minimal boson invariant mass.}
\end{table}

\begin{table}
\footnotesize
\begin{equation}
\begin{array}{|c||c|c|c|c|c|c|c|c|}
\hline
 \multicolumn{9}{|c|}{\mathrm{pp} \to V_1V_2(+\gamma) + X \mbox{ at }
   \sqrt{s} = 8 \mbox{ TeV}} \\
\hline
 \mbox{default cuts} & \multicolumn{2}{|c}{\PZ\PZ} &  \multicolumn{2}{|c}{\PW^+\PZ} &  \multicolumn{2}{|c}{\PW^-\PZ}& \multicolumn{2}{|c|}{\gamma\gamma}  \nonumber \\ 
p_{\rT}^{\cut}\,(\GeV) &  \sigma_\LO\,(\pba) & \delta_{\EW}(\%) &  \sigma_\LO\,(\pba) &
\delta_\EW(\%) &  \sigma_\LO\,(\pba) & \delta_\EW(\%) &  \sigma_\LO\,(\pba) & \delta_\EW (\%) \nonumber\\
\hline\hline
50  & 1.913 & -6.2  & 2.651 & -1.8 & 1.610 & -1.5 & 1.843 & \phantom{-}0.6 \nonumber\\
100 & 0.523 & -10.2 & 0.706 & -3.4 & 0.384 & -3.0 & 0.242 & -1.3 \nonumber\\
150 & 0.164 & -14.5 & 0.219 & -6.1 & 0.106 & -5.7 & 64.40\times
10^{-3} & -3.5 \nonumber\\
250 & 26.08\times 10^{-3} & -22.5 & 34.70\times 10^{-3} & -11.9 &
14.10\times 10^{-3} & -11.5 & 95.15\times 10^{-4} & -6.7 \nonumber\\
350 & 60.94\times 10^{-4} & -29.3 & 80.73\times 10^{-4} & -17.0 &
29.08\times 10^{-4} & -16.6 & 22.17\times 10^{-4} & -9.3 \nonumber\\
500 & 10.11\times 10^{-4} & -38.1 & 13.11\times 10^{-4} & -23.2 &
42.29\times 10^{-5} & -22.9 & 38.01\times 10^{-5} & -12.3 \nonumber\\
600 & 35.40\times 10^{-5} & -43.2 & 44.99\times 10^{-5} & -26.6 &
13.87\times 10^{-5} & -26.5 & 13.77\times 10^{-5} & -14.2 \nonumber\\
750 & 83.51\times 10^{-6} & -50.1 & 10.18\times 10^{-5} & -31.1 &
30.31\times 10^{-6} & -31.1 & 34.62\times 10^{-6} & -16.5 \nonumber\\
\hline
\end{array}
\end{equation}
\caption{\label{ta:totcs_ptcut_L8} Integrated leading-order cross sections
  and relative  EW corrections at LHC8 for different cuts on
  the minimal boson transverse momenta.}
\end{table}

\begin{table}
\footnotesize
\begin{equation}
\begin{array}{|c||c|c|c|c|c|c|c|c|}
\hline
 \multicolumn{9}{|c|}{\mathrm{pp} \to V_1V_2(+\gamma) + X \mbox{ at }
   \sqrt{s} = 8 \mbox{ TeV}} \\
\hline
\mbox{default cuts} & \multicolumn{2}{|c}{\PZ\PZ} &  \multicolumn{2}{|c}{\PW^+\PZ} &  \multicolumn{2}{|c}{\PW^-\PZ}& \multicolumn{2}{|c|}{\gamma\gamma}  \nonumber \\ 
M_{VV}^{\cut}\,(\GeV) &  \sigma_\LO\,(\pba) & \delta_{\EW}(\%) &  \sigma_\LO\,(\pba) &
\delta_\EW(\%) &  \sigma_\LO\,(\pba) & \delta_\EW(\%) &  \sigma_\LO\,(\pba) & \delta_\EW (\%) \nonumber\\
\hline\hline
200  & 3.271 & -4.9  & 4.728 & -1.5 & 2.980 & -1.4 & 0.579 & -0.3  \nonumber \\
300  & 0.950 & -7.1  & 1.921 & -2.5 & 1.117 & -2.2 & 0.169 & -1.6  \nonumber \\
400  & 0.360 & -8.7  & 0.821 & -3.5 & 0.439 & -3.2 & 66.45\times
10^{-3} & -2.6  \nonumber \\
500  & 0.159 & -10.2 & 0.387 & -4.5 & 0.191 & -4.2 & 30.52\times
10^{-3} & -3.6  \nonumber \\
700  & 39.70\times 10^{-3} & -13.5 & 0.103 & -6.9 & 44.69\times
10^{-3} & -6.5 & 84.71\times 10^{-4} & -5.2  \nonumber \\
800  & 21.27\times 10^{-3} & -15.4 & 56.47\times 10^{-3} & -8.3  &
23.07\times 10^{-3} & -7.8 & 48.66\times 10^{-4} & -5.8  \nonumber \\
1000 & 67.00\times 10^{-4} & -19.5 & 18.35\times 10^{-4} & -11.5 &
68.07\times 10^{-4} & -11.0 & 17.91\times 10^{-4} & -7.1  \nonumber \\
1200 & 23.88\times 10^{-4} & -23.7 & 68.65\times 10^{-4} & -14.4 &
23.82\times 10^{-4} & -14.0 & 72.84\times 10^{-5} & -8.3  \nonumber \\
1500 & 61.33\times 10^{-5} & -28.5 & 18.45\times 10^{-4} & -18.2 &
60.60\times 10^{-5} & -17.6 & 21.25\times 10^{-5} & -9.8  \nonumber \\
\hline
\end{array}
\end{equation}
\caption{\label{ta:totcs_mcut_L8} Integrated leading-order cross sections
  and relative  EW corrections at LHC8 for different cuts on the
  minimal boson invariant mass.}
\end{table}

\begin{figure}
\begin{center}
\includegraphics[width = 1.0\textwidth]{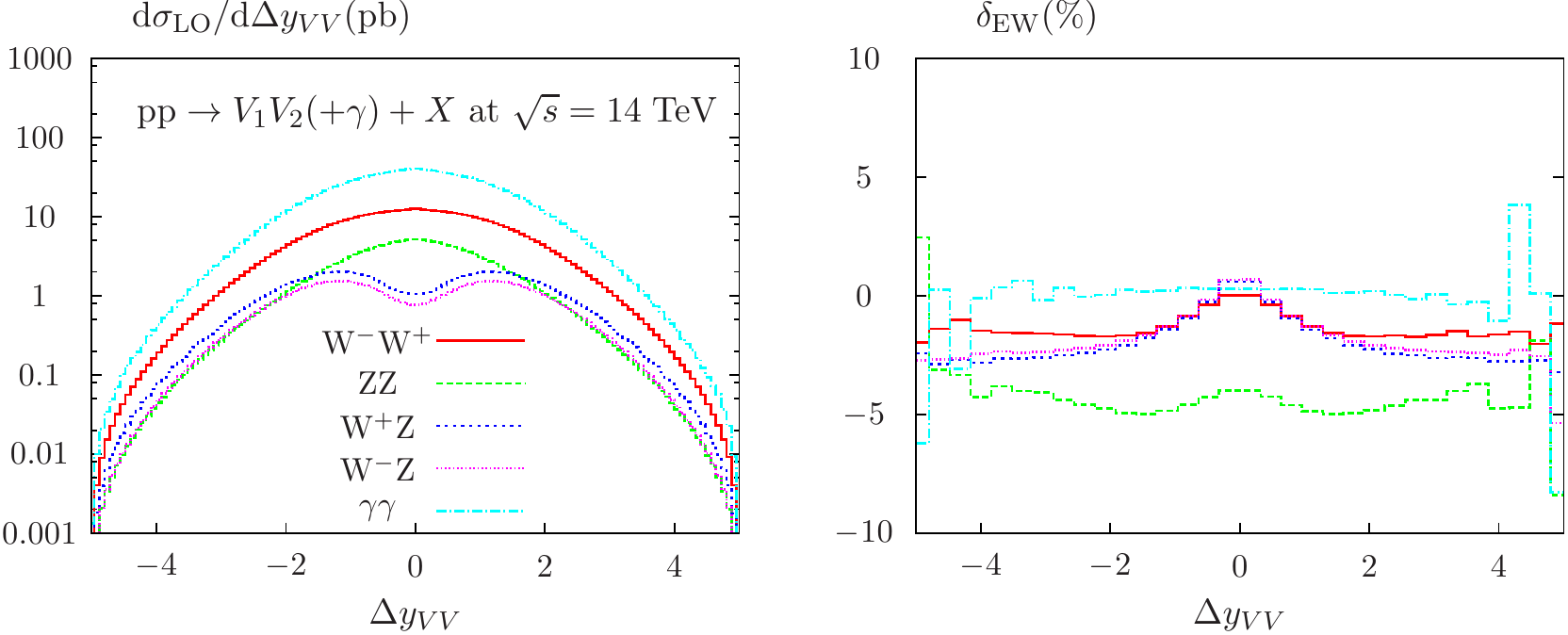}
\includegraphics[width = 1.0\textwidth]{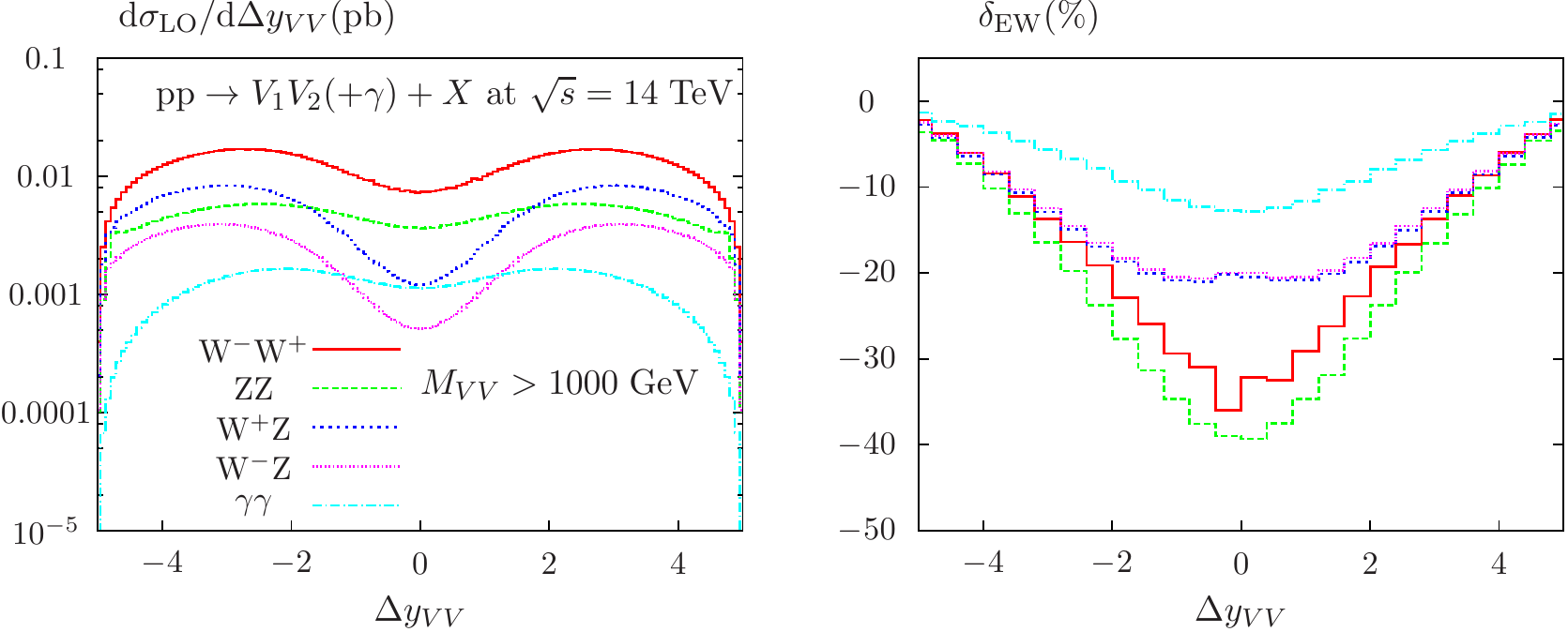}
\end{center}
\caption{\label{fi:dely_L14} Differential LO distributions of the boson
  rapidity gap (left) and corresponding EW corrections (right) at LHC14,
  evaluated with our default setup (top) and with a minimal boson
  invariant mass of 1000 GeV (bottom).}
\end{figure}

\begin{figure}
\begin{center}
\includegraphics[width = 1.0\textwidth]{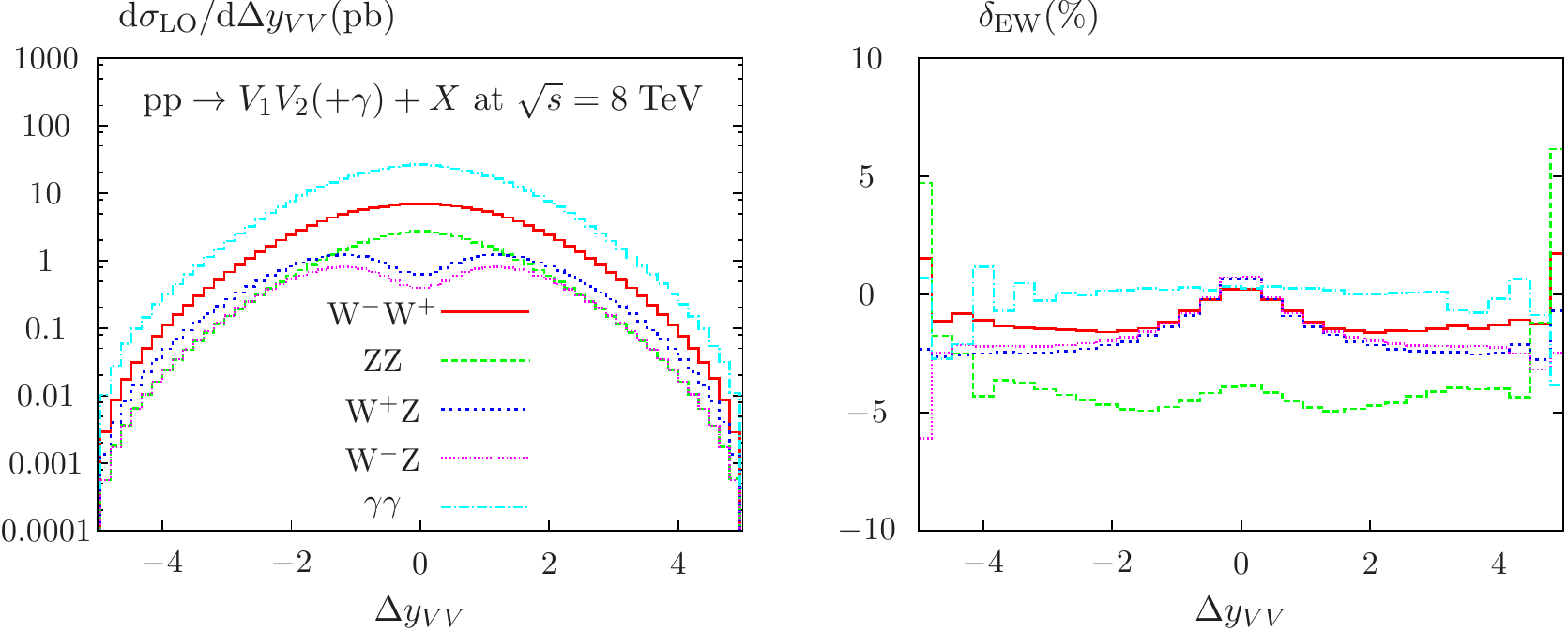}
\includegraphics[width = 1.0\textwidth]{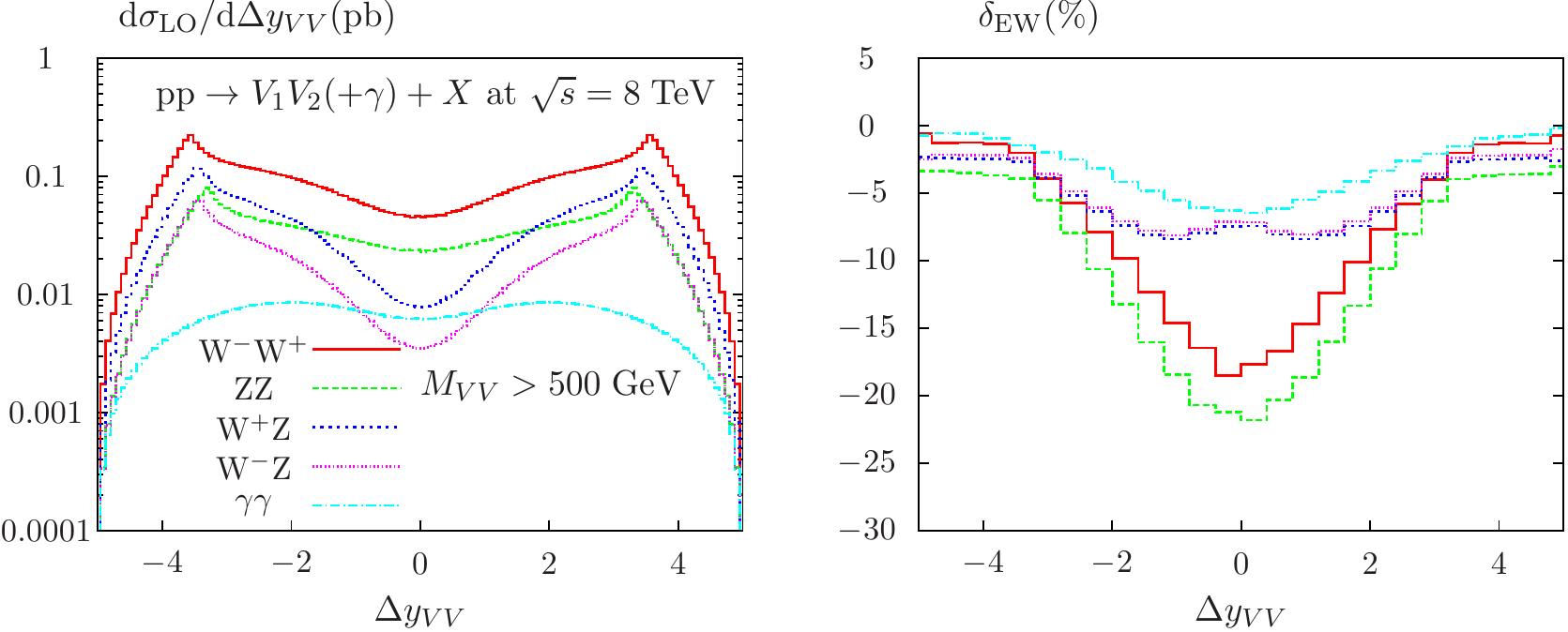}
\end{center}
\caption{\label{fi:dely_L8} Differential LO distributions of the boson rapidity gap (left)
  and corresponding EW corrections (right) at LHC8, evaluated with our
  default setup (top) and with a minimal boson invariant mass of 500 GeV
  (bottom).}
\end{figure}

Before discussing the EW corrections to $V$-pair production at
hadron colliders in detail let us first recall the most striking
features at leading order.

LO results for the processes $\mathrm{pp(\bar{p})} \to V_1 V_2 + X$ (see
eqs.~\eqref{eq:LO}) are given in table~\ref{ta:totcs_def} for integrated
cross sections evaluated with our default setup, in
table~\ref{ta:totcs_nocut} for the total cross sections without any
cuts. The relative EW corrections are also displayed and will be
discussed in detail in section~\ref{se:ew_corr}.  The rates for ZZ
and W$^-$Z production are in the same ballpark, both at the LHC
and the Tevatron. For the LHC, being a proton collider, the cross
section for W$^+$Z production is larger than for W$^-$Z production,
while at the Tevatron they trivially coincide. The WW
production cross sections, however, are roughly a factor of 5 (7) larger at the
LHC (Tevatron) than those for ZZ production. For our default cuts,
the cross section for $\gamma\gamma$ production is even larger
than the one for W-pair production, as a consequence of the
  singular behaviour for small $p_{\rT,\gamma}$.

  Requiring tighter cuts on the boson transverse momenta
  (tables~\ref{ta:totcs_ptcut} and~\ref{ta:totcs_ptcut_L8} and
  figure~\ref{fi:ptcut_vv}, left) and on invariant masses
  (tables~\ref{ta:totcs_mcut} and~\ref{ta:totcs_mcut_L8} and
  figure~\ref{fi:mcut_vv}, left), one observes a rapid decrease of the
  cross sections over several orders of magnitude, corresponding to the
  fact that $V$-pair production is dominated by both small scattering
  angles and low-energetic events, as has already been pointed out in
  ref.~\cite{Bierweiler:2012kw} for the WW case.  Consequently, a cut on
  the transverse momentum leads to a much stronger reduction than a
  comparable cut on the invariant mass, $M_{VV}^\cut = 2 p_{\rT}^\cut$,
  as is obvious from tables~\ref{ta:totcs_ptcut}
  and~\ref{ta:totcs_mcut}. In addition, also the relative rates of the
  different channels change dramatically. While comparable at low
  $p_{\rT}$, the cross section for $\PW^-\PZ$ production is 3 times
  smaller than the one for $\PW^+\PZ$ production when going to large
  values of $p_{\rT}$ and $M_{VV}$. For large $p_{\rT}$, ZZ production
  behaves similar to $\PW^+\PZ$ production, at large invariant masses
  the event rates are comparable to $\PW^-\PZ$ production. In
  contrast, the $\gamma\gamma$ cross section, which dominates at low
  transverse momenta, drops rapidly for high invariant masses, at high
transverse momenta it is comparable to $\PW^-\PZ$ production.

For high invariant masses the forward-backward peaking of the
cross sections is even more pronounced than for low energies as can be
deduced from the left plots of figures~\ref{fi:dely_L14}
and~\ref{fi:dely_L8}, where the differential cross sections are
presented as a function of the boson rapidity gap $\Delta y_{VV} =
y_{V_1}-y_{V_2}$, evaluated with our default cuts (top) and with a
minimal invariant mass of 1000~GeV (500 GeV) at LHC14 (LHC8)
(bottom). Note that $\Delta y_{VV} $ for fixed $M_{VV}$ corresponds to
the scattering angle in the diboson rest frame. At low
energies the distributions reach their maximum in the central region,
i.e.\ at rather low values of $|\Delta y_{VV}|$, corresponding to
central events, at high invariant masses they become maximal at
$|\Delta y_{VV}| \simeq 3$, corresponding to a drastic peaking in the
forward and backward directions.

\subsection{Electroweak corrections}
\label{se:ew_corr}
This section will be devoted to a detailed numerical analysis of the EW
corrections to vector-boson pair production at the LHC and Tevatron. The
corresponding effects on the total cross sections will be discussed in
section~\ref{se:ew_corr_tot}, those for partially integrated cross
sections in different kinematic regimes in
section~\ref{se:ew_corr_part}. In section~\ref{se:ew_corr_diff}, we
discuss the EW corrections to differential distributions of observables
relevant at the LHC, namely transverse momenta, rapidities and invariant
masses. In particular, we focus on the structure of the respective
corrections at highest energies, where high- as well as low-$p_\rT$
production of vector-boson pairs is addressed separately.

\subsubsection{Integrated cross sections}
\label{se:ew_corr_tot}
In a first step we present the EW corrections for the total
  production cross sections for WW, W$^\pm$Z, ZZ and $\gamma\gamma$ in
  table~\ref{ta:totcs_def}, using the default cuts described before. The
  EW corrections amount to roughly one percent, with the ZZ channel
  being the only exception with its sizable negative correction of about
  $-4\%$. As discussed below, this feature is present already close to
  production threshold, applies to Tevatron, LHC8 and LHC14 and is also
  present in rapidity distributions, as long as we do not enforce large
  $\hat{s}$ through selected cuts.

For reference purpose we also list the results for the total cross
section without any cuts for all final states apart of $\gamma\gamma$,
where the corresponding cross sections would diverge
(table~\ref{ta:totcs_nocut}). The size of the radiative corrections remains
practically unchanged.
  
\subsubsection{Partially integrated cross sections}
\label{se:ew_corr_part} 
\begin{figure}
\begin{center}
\includegraphics[width = 1.0\textwidth]{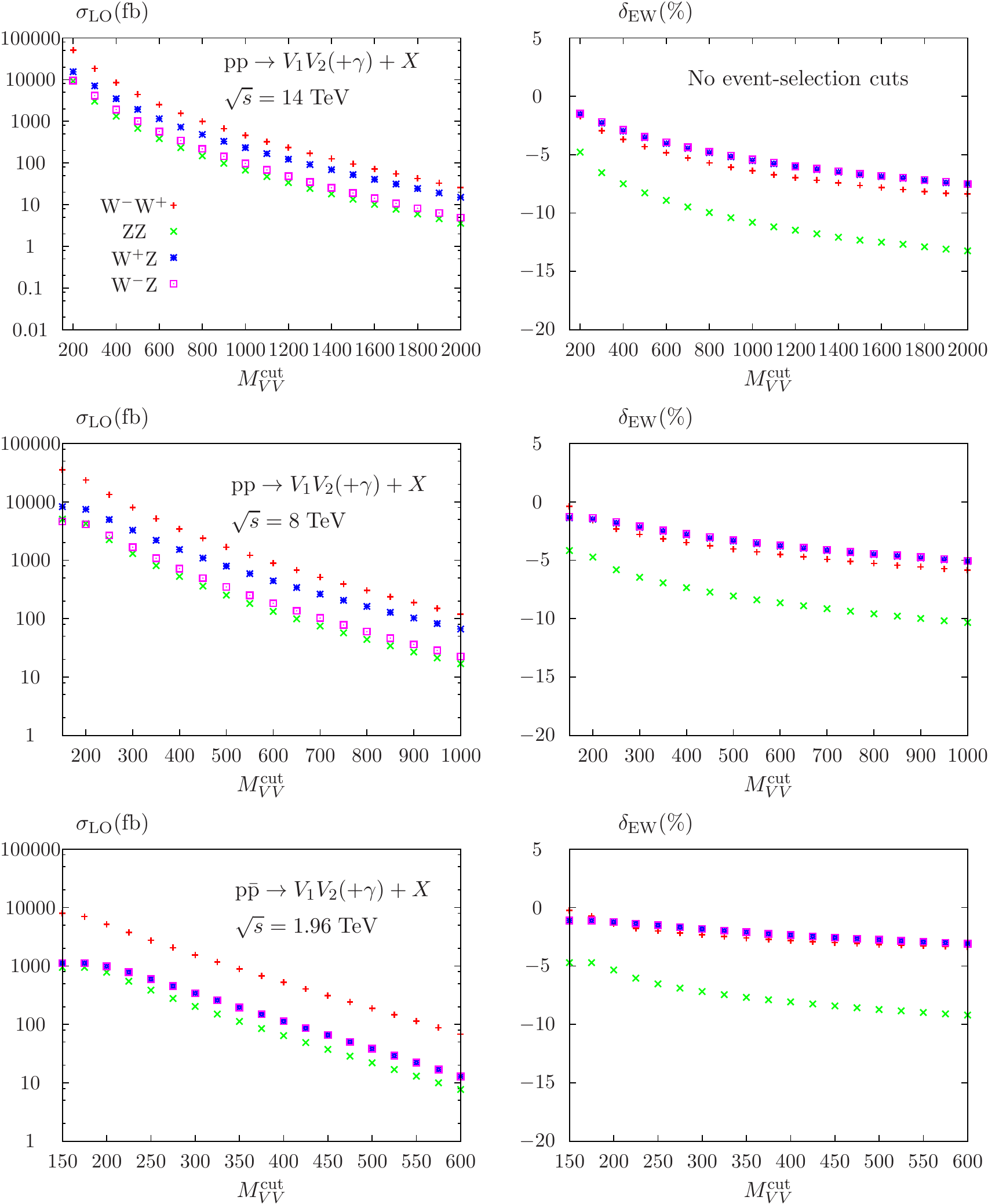}
\end{center}
\caption{\label{fi:mcut_vv_nocut}Integrated LO cross sections (left) and
  relative EW corrections (right) for different cuts on the invariant
 mass of the boson pair; no additional cuts on rapidity or transverse
  momentum are applied. The respective results are presented for LHC14
  (top), LHC8 (center) and the Tevatron (bottom).}
\end{figure}
In order to enhance hard scattering events and explore the TeV region it
is useful to consider partially integrated cross sections, where
additional cuts are introduced on the transverse momenta of the gauge
bosons or on the invariant mass of the gauge-boson pair. Numerical
results are listed in
tables~\ref{ta:totcs_ptcut},~\ref{ta:totcs_mcut},~\ref{ta:totcs_ptcut_L8}
and~\ref{ta:totcs_mcut_L8} and in figures~\ref{fi:ptcut_vv}
and~\ref{fi:mcut_vv}.

As expected, one observes increasingly negative corrections with
increasing $p_{\rT}^\cut$ or $M_{VV}^\cut$. The corrections are
largest for ZZ production, reaching -50\% at LHC14 for $p_\rT^\cut =
800$~GeV, and smallest for $\gamma\gamma$ production, where they remain
below 20\% for all kinematic configurations. As already stated in our
paper on WW production, the EW corrections at the LHC are considerably
more important than at the Tevatron. 

For illustration we also present the results up to $p_{\rT}^{\cut} =
1$~TeV which might be accessible in a high-luminosity run. In this case
corrections exceeding $40\%$ are observed in W-pair production and the
question of two-loop contributions necessarily arises. For W-pair
production the logarithmically enhanced (NNLL) corrections have been
discussed in ref.~\cite{Kuhn:2011mh}, for the other final states they
are not yet available and we will not dwell further on this subject. 

The particularly large negative EW corrections in case of Z-pair
production may be understood as a consequence of two completely
independent physical effects. On the one hand, the EW corrections to the
ZZ cross section exhibit a constant offset of $-4\%$ irrespective of
cuts when compared to WW, WZ and $\gamma\gamma$ production. This feature
is most obvious in figure~\ref{fi:mcut_vv_nocut}, where the
$M_{VV}^\cut$ dependence of the cross sections is displayed without
additional rapidity or transverse-momentum restrictions, i.e.\ without
the default cuts described in section~\ref{se:setup}.  Sudakov-enhanced
logarithms cannot be made responsible, since a $-4\%$ correction is
already present in the threshold region, and it is hardly dependent on
the actual value of the collider energy. We additionally observe that
real radiation of hard photons, which is included in our results
together with radiation of soft and virtual photons, is particularly
small for ZZ production (see figure~\ref{fi:real_L14}, bottom). Taking
both effects into account, the EW corrections at high $p_{\rT}$ which
can be attributed to large logarithms stemming from the weak corrections
are similar for WW and ZZ production, while they are significantly
smaller in case of $\PW^\pm\PZ$ production.

Nevertheless, also corrections to W$^-$Z production are sizable,
reaching $-35\%$ at LHC14 for $p_{\rT}^\cut = 800$ GeV. Furthermore they
are quite similar to those for W$^+$Z production, since at parton level
the corresponding unpolarized cross sections coincide. Small deviations
solely arise from the different parton-luminosities multiplying
the real-radiation and virtual contributions, respectively, leading to
slightly different relative corrections, since the differential partonic
corrections $\hat{\delta}_{\EW}(\hat{s},\hat{t})$, which evidently are
the same for both channels, enter the hadronic results with different
weights.

Going back to figure~\ref{fi:mcut_vv_nocut}, we find that the relative
corrections are small (below 10\%) even for large invariant masses
(without additional cuts on rapidities and transverse momenta), since
  vector-boson pair production is dominated by low scattering angles,
  i.e.\ small $|\hat{t}|$, where no logarithmically enhanced EW
  corrections are expected. Surprisingly enough, even moderate cuts on the boson
transverse momenta and rapidities lead to considerable relative
corrections as shown in figure~\ref{fi:mcut_vv}.

Another interesting finding is that the EW corrections in case of
$\gamma\gamma$ production are again substantially smaller than in the
massive channels, reaching $-20\%$ for $p_{\rT}^{\cut} = 1000$~GeV,
  which is consistent with corresponding results from
  ref.~\cite{Layssac:2001ur}, where logarithmic corrections to the
  processes $\gamma\gamma \to f\bar{f}$ were computed in the
  high-$p_{\rT}$ approximation.\footnote{The full one-loop EW
    corrections to $\gamma\gamma \to \mathrm{t\bar{t}}$  were
    presented even earlier in ref.~\cite{Denner:1995ar}.} As already
  stated above, up to now no phenomenological study at all exists of
  the EW effects in photon-pair production at hadron colliders.

\subsubsection{Differential distributions}
\label{se:ew_corr_diff} 
\begin{figure}
\begin{center}
\includegraphics[width = 1.0\textwidth]{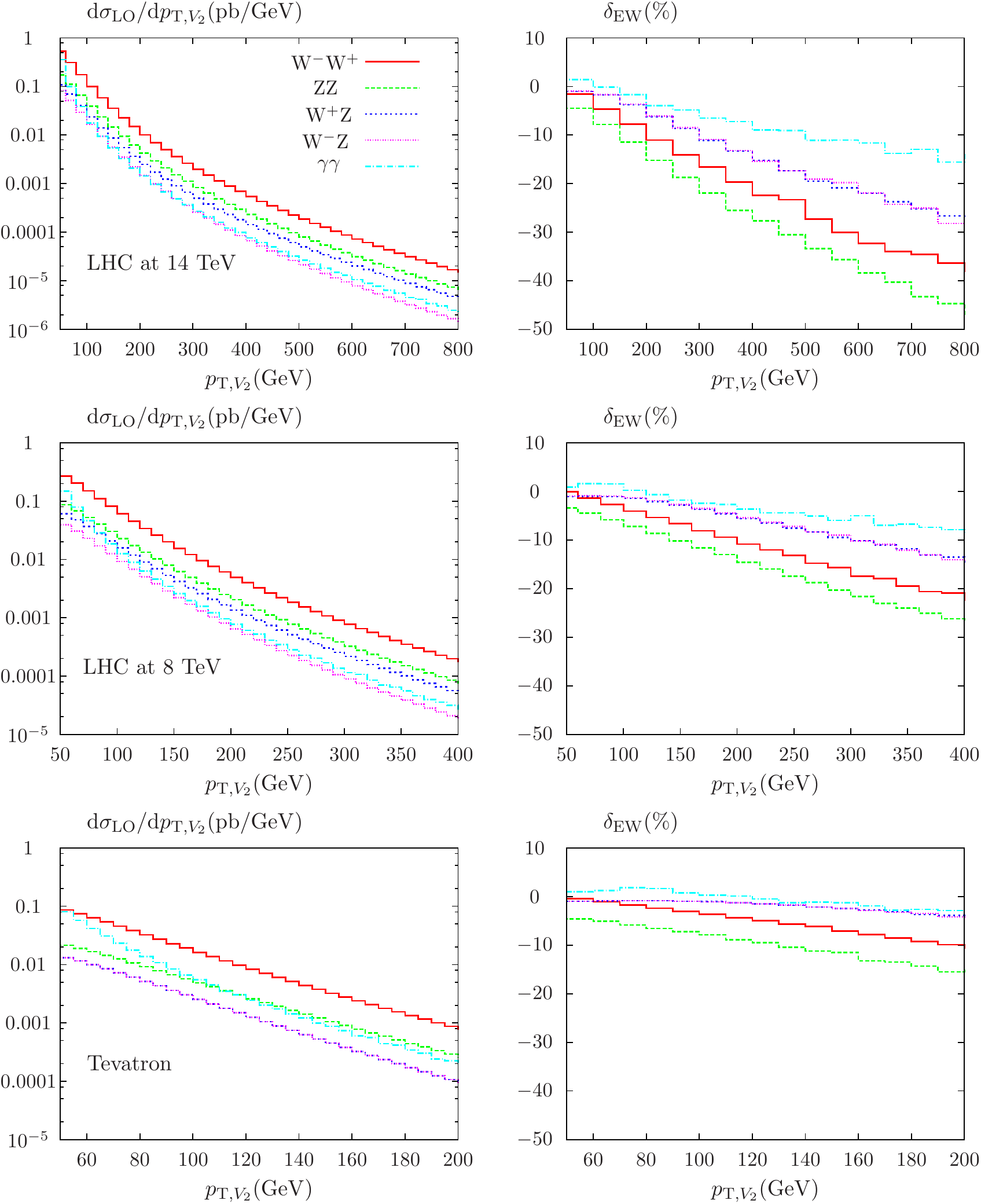}
\end{center}
\caption{\label{fi:dsdpt_all} Differential LO distributions of the transverse momentum (left)
  and corresponding relative EW corrections
  (right) evaluated with our default setup. The respective results are
  presented for LHC14 (top), LHC8 (center) and the Tevatron (bottom).}
\end{figure}
\begin{figure}
\begin{center}
\includegraphics[width = 1.0\textwidth]{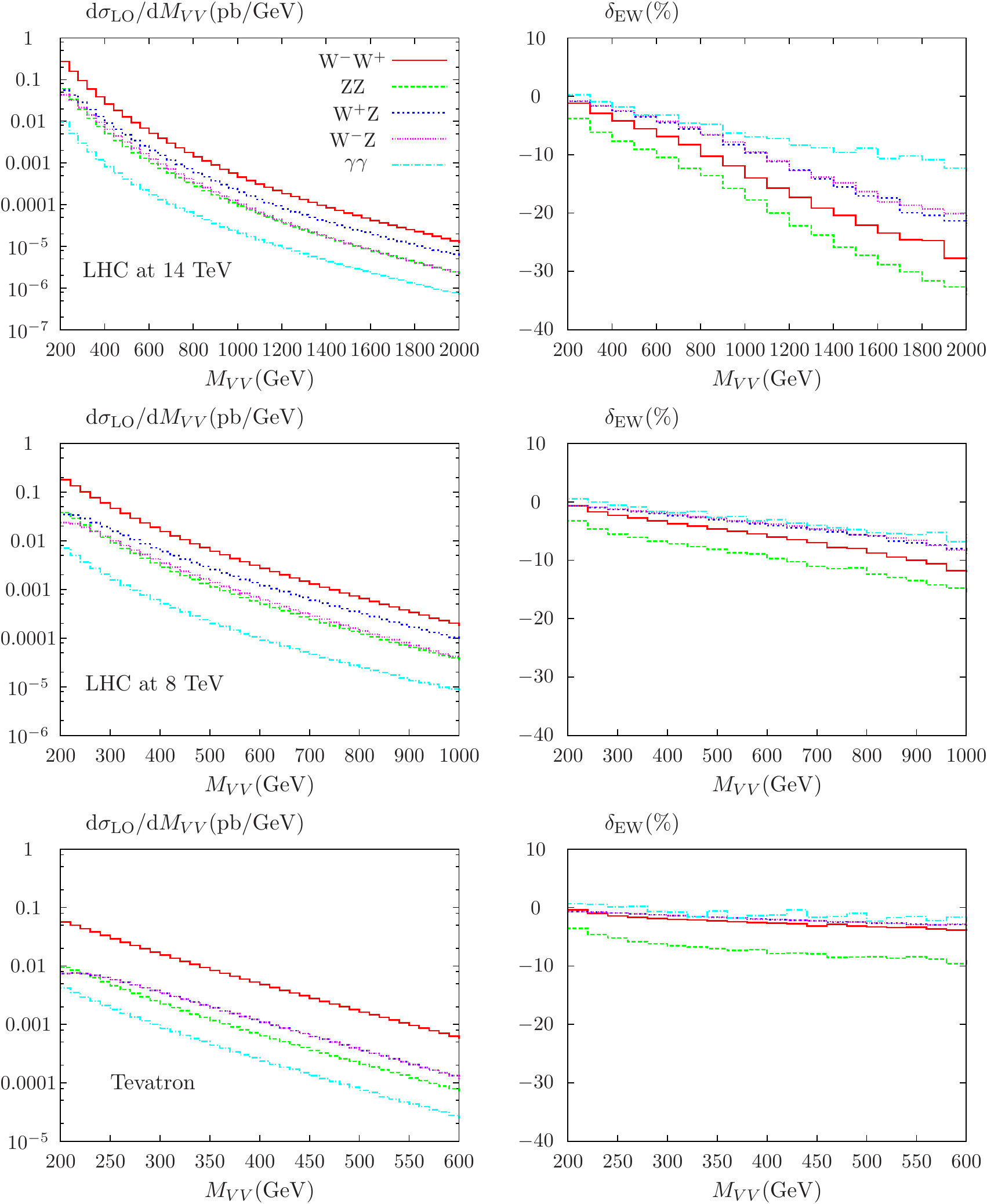}
\end{center}
\caption{\label{fi:dsdm_all} Differential LO distributions of the invariant mass (left) and corresponding relative EW corrections
  (right) evaluated with our default setup. The respective results are
  presented for LHC14 (top), LHC8 (center) and the Tevatron (bottom).}
\end{figure}
\begin{figure}
\begin{center}
\includegraphics[width = 1.0\textwidth]{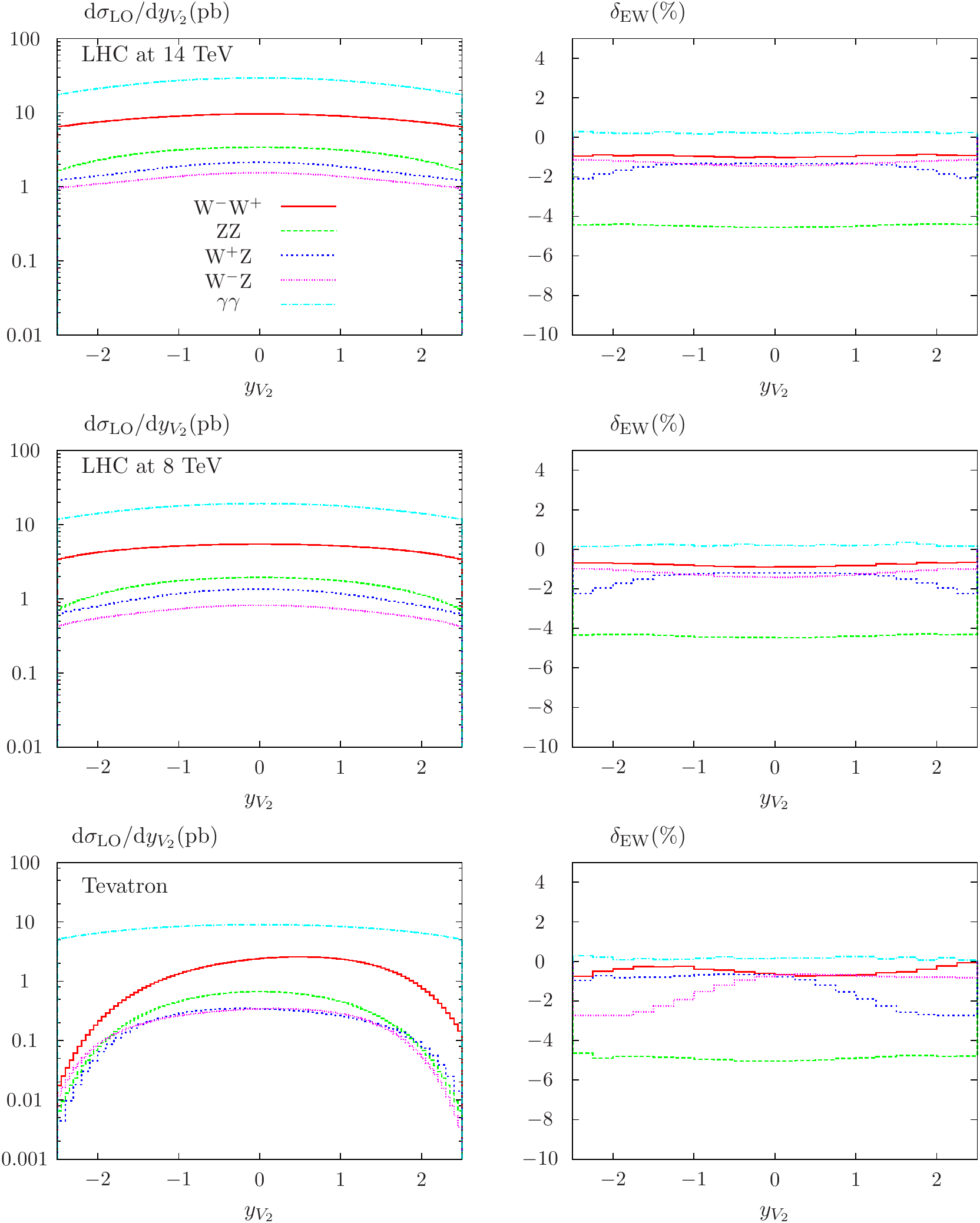}
\end{center}
\caption{\label{fi:dsdy_all} Differential LO distributions of the rapidity (left) and corresponding  relative EW corrections
  (right) evaluated with our default setup. The respective results are
  presented for LHC14 (top), LHC8 (center) and the Tevatron (bottom).}
\end{figure}

\begin{figure}
\begin{center}
\includegraphics[width = 1.0\textwidth]{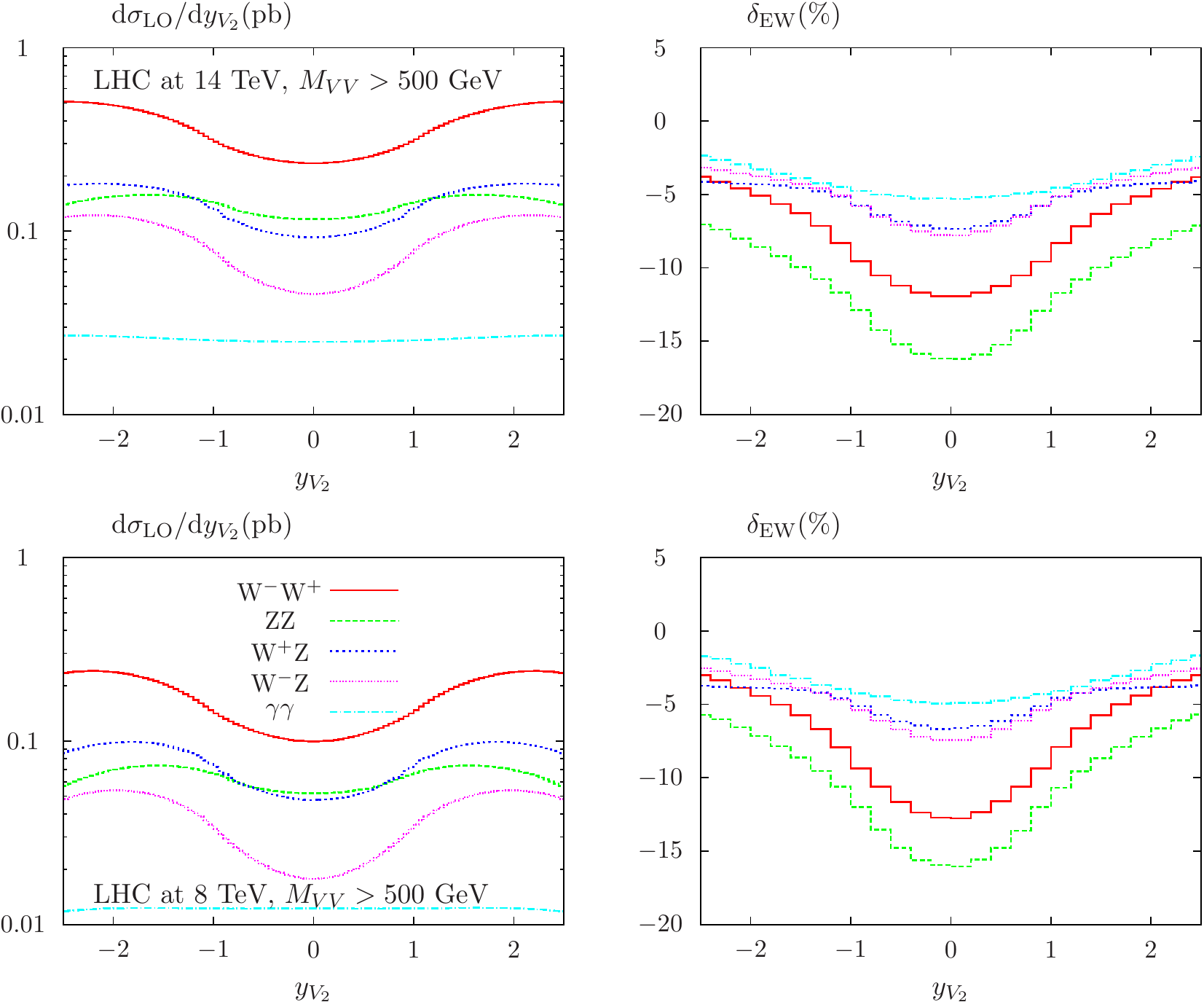}
\end{center}
\caption{\label{fi:dsdy_500}Differential LO distributions of the boson rapidity (left) and
  corresponding EW corrections (right) at LHC14 (top) and LHC8 (bottom)
  for a minimal invariant mass of 500 GeV.}
\end{figure}

\begin{figure}
\begin{center}
\includegraphics[width = 1.0\textwidth]{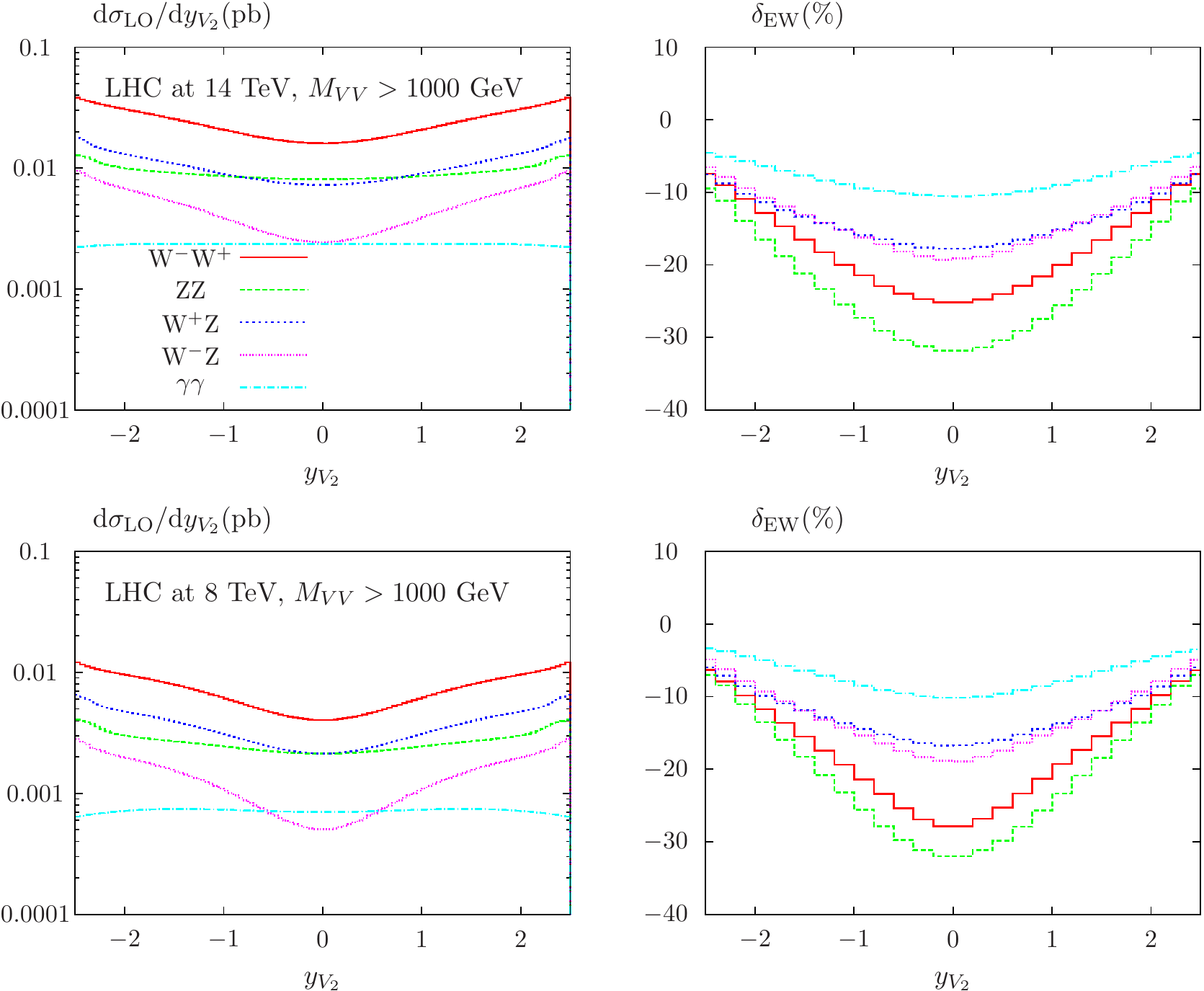}
\end{center}
\caption{\label{fi:dsdy_1000}Differential LO distributions of the boson rapidity (left) and
  corresponding EW corrections (right) at LHC14 (top) and LHC8 (bottom)
  for a minimal invariant mass of 1000 GeV.}
\end{figure}

\begin{figure}
\begin{center}
\includegraphics[width = 1.0\textwidth]{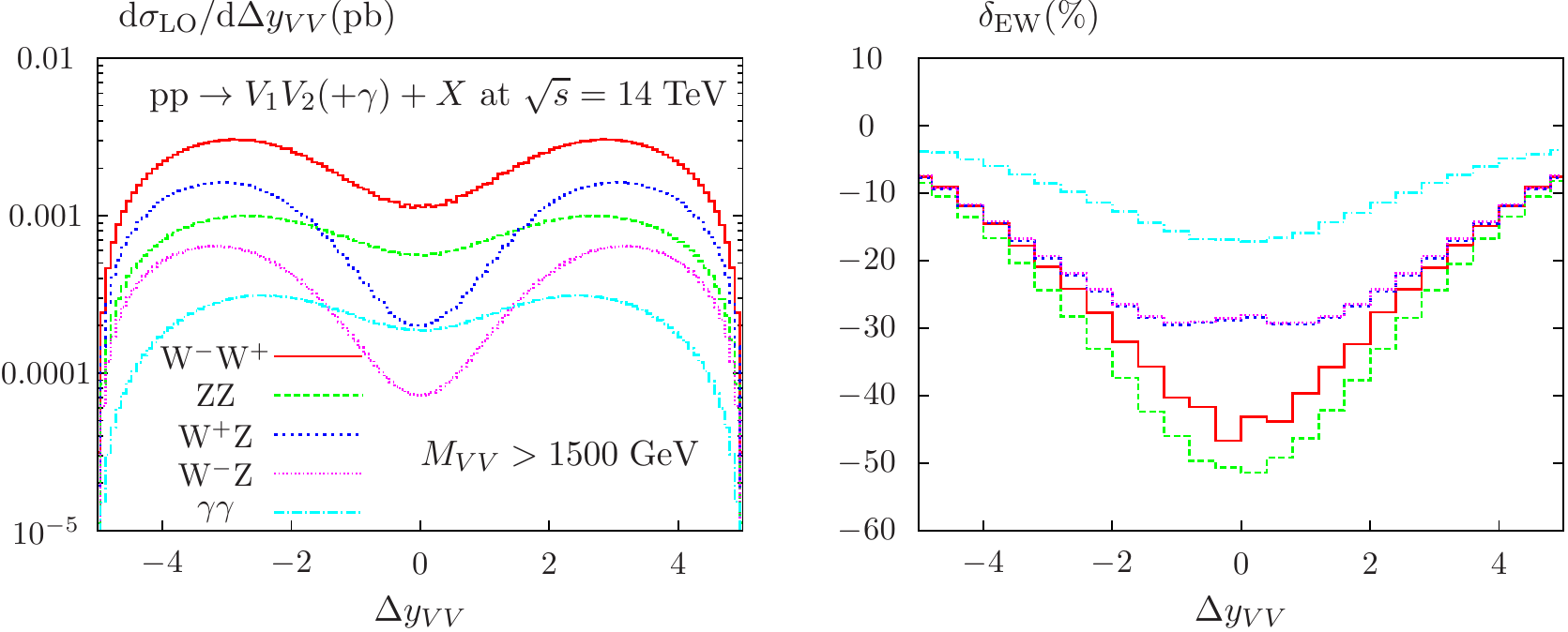}
\includegraphics[width = 1.0\textwidth]{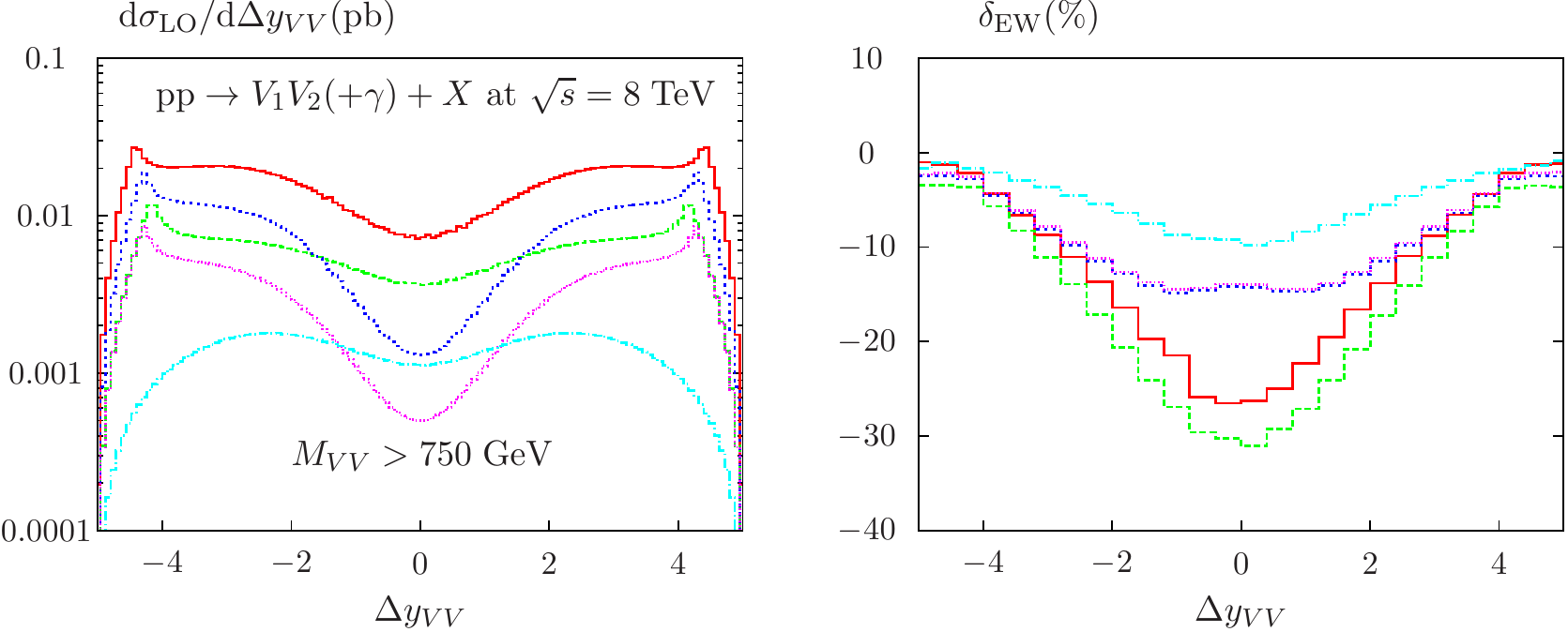}
\end{center}
\caption{\label{fi:dely_high} Differential LO distributions of the boson
  rapidity gap (left) and corresponding EW corrections (right) for a
  minimal invariant mass of 1500 GeV at LHC14 (top) and 750 GeV at LHC8
  (bottom), respectively.}
\end{figure}

In this section differential cross sections for various kinematic
scenarios are investigated. Specifically, in addition to transverse-momentum,
invariant-mass and rapidity distributions generated in our default
setup, we also present results with explicit cuts on the invariant mass
of the vector bosons. This allows to investigate the EW corrections in the
high-energy regime where new-physics signatures might have a
sizable impact. As will be shown, significant distortions of the
angular distributions are observed which might easily be misinterpreted
as a signal of anomalous couplings and hence new physics.  

In figures~\ref{fi:dsdpt_all}, \ref{fi:dsdm_all} and~\ref{fi:dsdy_all}
we display the differential distributions and corresponding EW
corrections for the boson transverse-momentum, invariant-mass and
rapidity distributions, respectively, evaluated in the default setup
defined in section~\ref{se:setup}. As already stated for the partially
integrated cross sections in the previous section, the LO distributions
rapidly decrease with increasing values of $p_{\rT}$ and $M_{VV}$,
reflecting that vector-boson pair production is in general dominated by
events with low $\hat{s}$, as a consequence of the rapidly-falling PDFs
and cross sections. At the same time, the EW corrections increase with
$p_\rT$ and $M_{VV}$, ranging between $-15\%$ for $\gamma\gamma$ and
$-45\%$ for ZZ if we consider $p_{\rT}$ values of 800 GeV at LHC14 as
example.

The relative corrections for the rapidity distributions
(figure~\ref{fi:dsdy_all}, right) are small for WW, WZ and $\gamma\gamma$
production and resemble the corrections for the integrated cross sections
presented in table~\ref{ta:totcs_def}. Again, for ZZ production a
constant offset of $-4\%$ in the EW corrections is evident, reflecting a
remarkably constant $K$-factor. This behaviour is expected, since
  both the total cross section and the rapidity distributions are
  dominated by small $M_{VV}$.

  Let us now consider vector-boson pair production at highest energies
  accessible at the LHC. This is achieved by restricting $M_{VV}$ to
  values above 500 GeV (fig.~\ref{fi:dsdy_500}) and 1~TeV
  (fig.~\ref{fi:dsdy_1000}).  Looking at the right panels of
  figure~\ref{fi:dsdy_500} and figure~\ref{fi:dsdy_1000}, we observe
  that the relative EW corrections to the boson rapidity distributions
  are sizable at low rapidities, corresponding to central events with
  high transverse momenta, i.e.\ the Sudakov region. Large rapidities,
  in contrast, correspond to small scattering angles (small $|\hat{t}|$
  or $|\hat{u}|$) and thus small $p_{\rT}$. EW corrections per se are
  logarithmically enhanced only in the Sudakov region. For identical
  cuts on the invariant mass $M_{VV}$ the corrections are very similar
  at LHC14 and LHC8, respectively, reaching $-15\%$ ($-30\%$) for
  $M_{VV} > 500$ GeV ($M_{VV} > 1000$ GeV). The corresponding
  corrections for the distributions with respect to $\Delta y_{VV} =
  y_{V_1} - y_{V_2}$ were presented in figures~\ref{fi:dely_L14}
  and~\ref{fi:dely_L8} for $M_{VV} > 500$ GeV ($M_{VV} > 1000$ GeV) for
  LHC8 (LHC14). Large corrections were observed for small rapidity gaps,
  amounting to nearly $-40\%$ ($-25\%$) for central ZZ production at
  LHC14 (LHC8).

    This behaviour can be pushed even further by going to $M_{VV}> 1.5$
    TeV, a region potentially accessible at a high luminosity
    LHC. Results for a minimal invariant mass of $M_{VV} > 750$ GeV
    ($M_{VV} > 1500$ GeV) at LHC8 (LHC14) are shown in
    figure~\ref{fi:dely_high}, and for Z-pair production at LHC14 in
    this kinematic region the corresponding corrections reach
    $-50\%$. In this regime, also weak two-loop effects might be required
    to reliably predict the cross sections.

\subsubsection{Comparison with existing results}
To additionally validate our numerical analysis and to assess the
remaining theoretical uncertainties, we compare our predictions for
vector-boson pair production at hadron colliders with older results
obtained in a high-energy
approximation~\cite{Accomando:2004de}. Although a tuned comparison in
general is not possible since the approach taken in the present work
does not allow to apply event-selection cuts to the leptonic decay
products of the vector bosons, as is done in
ref.~\cite{Accomando:2004de}, (See, however, the discussion in
section~\ref{se:weakZZ}.), qualitative statements can be made already
now.

Comparing our results for Z-pair production
($\delta_{\mathrm{EW}}^{\mathrm{ZZ}}$) to those given in Table 3 of
ref.~\cite{Accomando:2004de} ($\delta_{\mathrm{EW}}^{\mathrm{ZZ,
    ADK}}$), we observe very good agreement in the whole energy range
considered if we employ the additional constraint $|\Delta y_{\PZ\PZ}| <
3$ on the rapidity gap of the Z bosons to explicitly enforce Sudakov
kinematics (i.e.\ $\hat{s},|\hat{t}|,|\hat{u}| \gg M_{\PZ}^2$). For
instance, we find $\delta_{\mathrm{EW}}^{\mathrm{ZZ}} = -28.5\%$, to be
compared with $\delta_{\mathrm{EW}}^{\mathrm{ZZ, ADK}} = -28.1\%$ for
$M_{\PZ\PZ}>1000$ GeV. As expected, EW corrections to ZZ production in
the Sudakov regime are exhaustively described by logarithmic weak
corrections, and mass effects do not play a significant
role. Surprisingly, also off-shell effects and final-state photon
radiation, both included in Ref.~\cite{Accomando:2004de}, as well as LHC
acceptance cuts, do not seem to noticeably affect the relative EW
corrections after the Z reconstruction has been performed.

Turning to $\PW\PZ$ production, the comparison is also
  straightforward. According to scenario (7.3) of
  ref.~\cite{Accomando:2004de} we apply--in addition to our default
  cuts--a cut on the transverse momentum of the Z boson and obtain
  $\delta_{\EW}^{\PW^+\PZ} = -22.4\%$ for $p_{\rT,\PZ} > 500$~GeV, which
  is in reasonable agreement with $\delta_{\mathrm{EW}}^{\mathrm{WZ,
      ADK}} = -21.2\%$ given in table 1 of
  ref.~\cite{Accomando:2004de}.


\subsection{Z-pair production: polarization and decays}
\label{se:weakZZ}

Once sufficiently large samples of gauge-boson pairs have been produced,
it will be important to investigate the angular distributions of their
decay products, which evidently carry the information about the Z (or W)
polarization. 

Up to the order considered in the present paper, real plus virtual
photon radiation can be separated in a gauge-invariant manner, as far as
ZZ production is concerned. It is quite remarkable that this purely
electromagnetic subset of the corrections to ZZ production is tiny, in
general below $1\%$ in all cases discussed in this paper.  In a first
step we thus evaluate the cross sections for the production of polarized
Z pairs and the impact of purely weak corrections on these cross
sections. In a second step, we consider combined production and decay
for the mode $\mathrm{pp} \to \PZ(\to \Pe^+\Pe^-)\,\PZ(\to
\mu^+\mu^-)+X$, including weak corrections.

\subsubsection{Polarization effects}
Let us start with polarized Z-pair production at LHC8 and
LHC14. Longitudinal, right- and left-circular production are denoted by
(L), $(+)$ and $(-)$. The cross sections for the production of one Z
with polarization $(i)$ and one with polarization $(j)$ is represented by
$\sigma(ij)$. The unpolarized cross section is thus composed of the
following combination
\begin{equation}
\sigma_{\mathrm{tot}} = \sigma(\rL\rL)+ \sigma(++)+ \sigma(--)+ \sigma(+-)+
\sigma(\rL+)+ \sigma(\rL-)\,. 
\end{equation}
As a consequence of CP symmetry $\sigma(\rL+) = \sigma(\rL-)$ and
$\sigma(++) = \sigma(--)$. The remaining four independent
combinations are listed in table~\ref{ta:pol_ZZ}. The results are
presented for the default cuts, for events with $p_{\mathrm{T,Z}} >
500$~GeV and for $p_{\mathrm{T,Z}} > 1000$~GeV. 

Let us first discuss the Born cross section, which is the upper entry
for each partial or summed cross section. Already for the default cuts
it receives its major contribution (70\%) from the ($+-$) configuration,
for larger transverse momenta the remaining configurations die out
quickly. This behaviour can be deduced directly from the equivalence
theorem: neutral scalar pair production is strictly forbidden, which in
the present case leads to the $M_{\PZ}^2/\hat{s}$ suppression of
$\sigma(\mathrm{\rL\rL})$. Also the ($++$) configuration is strictly
forbidden for massless gauge bosons. The $M_{\PZ}/\sqrt{\hat{s}}$
behaviour of $\sigma(\rL\pm)$ is similar to the one for W pairs e.g.\
discussed in ref.~\cite{Beenacker:1991uc}. Qualitatively, this behaviour
can indeed be read off from a comparison of the ratios
$\sigma(\rL\rL)/\sigma(+-)$ and $\sigma(\rL\pm)/\sigma(+-)$ for the
$p_{\mathrm{T}}^{\mathrm{cut}}$ values of 500~GeV and 1000~GeV.

Also shown in table~\ref{ta:pol_ZZ} as lower entry are the $\O (\alpha)$
weak corrections $\delta\sigma_{\mathrm{weak}}$. Again we observe the
overall reduction by about 5\%, arising mainly from small $\hat{s}$,
which is fairly similar for the different polarizations, as far as low
$\hat{s}$ are concerned. For larger $p_{\rT}$ values a different pattern
emerges. For the suppressed diagonal configurations (LL), $(++)$, $(--)$
the negative corrections increase with $p_{\rT}$, and quickly exceed the
Born contribution. Hence, if one would try to analyze the different
polarizations separately, one should include the squared 1-loop
correction. The corresponding values for $\delta\sigma_{\mathrm{weak}}$
representing the interference between Born and one-loop term plus
one-loop squared contribution are given in round brackets in the same
line. However, since all these contributions are below one permille,
they are irrelevant for all practical considerations. Note, in addition,
that the residual uncertainties on the integrated cross sections due to
missing higher-order weak corrections are at the level of 10\% and 20\%
for $p_{\rT,\PZ}>500$~GeV and $p_{\rT,\PZ}>1000$~GeV, respectively.

In total a fairly simple pattern emerges: for small $p_{\rT}$
electroweak corrections are very similar for all polarizations and can
be taken as one global factor, for large $p_{\rT}$ only one combination
survives and corrections are again trivially represented by one
factor. Electroweak corrections can therefore be represented for the
bulk of events at each $\hat{s}$ and $\hat{t}$ by a correction factor
which does not modify the relative importance of the different
polarizations.
\begin{table}
\footnotesize
\begin{equation}
\begin{array}{|c||c|c|c|c|c|}
\hline
 \multicolumn{6}{|c|}{\mathrm{pp} \to \mathrm{ZZ} + X} \\
\hline
\mbox{ZZ polarizations} & \mbox{summed} &\mathrm{LL} & \mathrm{L+}& ++ &+-  \nonumber \\
\hline
\hline 
\multicolumn{6}{|c|}{\mbox{LHC14}} \\
\hline
\mbox{default cuts} &&&&& \nonumber \\
\sigma_{\mathrm{LO}}/\mathrm{pb}&7.067&0.402&0.734&0.100 &4.997 \nonumber \\
\delta \sigma_{\mathrm{weak}}/\mathrm{pb}&-0.338(-0.292)&-0.015(-0.014)&-0.029(-0.025)&-0.004(-0.003)&-0.257(-0.223) \nonumber \\
\hline
p_{\mathrm{T,Z}}>500\;\GeV &&&&& \nonumber \\
\sigma_{\mathrm{LO}}/\mathrm{pb}&10^{-2} \times [0.499 &10^{-7} \times
[0.921&10^{-4} \times[ 0.334&10^{-7} \times [0.230&10^{-2} \times [0.492\nonumber \\
\delta\sigma_{\mathrm{weak}}/\mathrm{pb}&-0.195(-0.148)]&-4.70(+5.577)]&-0.087(-0.067)]&-0.426(-0.185)]&-0.192(-0.147) ]\nonumber \\
\hline
p_{\mathrm{T,Z}}>1000\;\GeV&&&&& \nonumber \\
\sigma_{\mathrm{LO}}/\mathrm{pb}&10^{-3} \times [0.146 & 10^{-9} \times
[0.189 & 10^{-6} \times [ 0.306 & 10^{-10} \times [0.475&10^{-3} \times [0.146\nonumber \\
\delta\sigma_{\mathrm{weak}}/\mathrm{pb}&-0.088(-0.062)]&-4.319(+30.04)]&-0.126(-0.090)]&-2.953(+2.295)]&-0.088(-0.062)]\nonumber \\
\hline\multicolumn{6}{|c|}{\mbox{LHC8}} \\
\hline
\sigma_{\mathrm{LO}}/\mathrm{pb}&3.810&0.223&0.396&10^{-1}\times[0.559&2.676 \nonumber \\
\delta \sigma_{\mathrm{weak}}/\mathrm{pb}&-0.179(-0.155)&-0.009(-0.008)&-0.016(-0.014)&-0.002(-0.002)]&-0.134(-0.117) \nonumber \\
\hline
p_{\mathrm{T,Z}}>500\;\GeV &&&&& \nonumber \\
\sigma_{\mathrm{LO}}/\mathrm{pb}&10^{-2} \times [0.101&10^{-7}\times
[0.202&10^{-5} \times [0.779&10^{-8}\times[0.504&10^{-3}\times[0.996\nonumber \\
\delta\sigma_{\mathrm{weak}}/\mathrm{pb}&-0.039(-0.030)]&-0.975(+0.748)]&-0.204(-0.157)]&-0.895(-0.425)]&-0.383(-0.293)] \nonumber \\
\hline
p_{\mathrm{T,Z}}>1000\;\GeV&&&&& \nonumber \\
\sigma_{\mathrm{LO}}/\mathrm{pb}&10^{-5}\times[0.919&10^{-10}\times[0.121&10^{-7}\times[0.231&10^{-11}\times[0.303&10^{-5}\times[0.915\nonumber \\
\delta\sigma_{\mathrm{weak}}/\mathrm{pb}&-0.557(-0.387)]&-2.599(+14.909)]&-0.098(-0.070)]&-1.742(+1.043)]&-0.555(-0.387)]\nonumber \\
\hline
\end{array}
\end{equation}
\caption{\label{ta:pol_ZZ} Polarized LO cross sections and corresponding
  weak corrections to ZZ production at the LHC for different cuts on the
  boson transverse momenta. The first entry for the corrections represents
  the interference between Born and one-loop amplitude, the second entry
  (in brackets)
  includes the squared one-loop amplitude.}
\end{table}
\subsubsection{Leptonic decays}
Let us, in a next step, evaluate the complete production and decay
process for the $\Pe^+\Pe^-\mu^+\mu^-$ final state at the LHC. For the
acceptance cuts on muon and electron transverse momenta and rapidities
we adopt the prescriptions of ref.~\cite{Accomando:2004de}, namely
\begin{equation}
  p_{\rT,l} > 15\;\mathrm{GeV}\,,\quad |y_l|< 3\,.
\end{equation}
The intermediate Z bosons are reconstructed from the final-state
leptons requiring 
\begin{equation}
|M_{l\bar{l}} - M_{\PZ}| < 20\;\mathrm{GeV}
\end{equation}
to suppress the admixture of virtual photons and to improve the validity
of the approximations used to compute the weak corrections.

The cuts employed in table~\ref{ta:zz4l} are chosen to mimic on the one
hand the experimental acceptance, and on the other hand, select events
with increasing $\sqrt{\hat{s}}$, corresponding to the invariant mass
$M_{\mathrm{inv}}(4l)$ of the four-lepton system. The first four columns
represent predictions for LHC14 and LHC8 using four variants of the Born
approximation. In the first column we give the full LO cross section,
including all off-shell effects and non-resonant contributions, in a
naive fixed-width implementation using non-vanishing constant
vector-boson widths\footnote{For the numerical evaluation we adopt the
  values from ref.~\cite{Accomando:2004de} for $\Gamma_\PZ$ and $\Gamma_\PW$.}
\begin{equation}
\Gamma_{\PZ} = 2.505044\;\mathrm{GeV}\,,\quad \Gamma_{\PW} = 2.099360\;\mathrm{GeV}\,
\end{equation}
in propagators  with time-like
4-momenta,
\begin{equation}
\frac{1}{p^2-M_V^2} \to \frac{1}{p^2-M_V^2+\theta(p^2)\ri M_V\Gamma_V}\,.
\end{equation}
The second column shows the corresponding results evaluated
in the complex-mass scheme (CMS)~\cite{Denner:2005fg} which involves
complex couplings. In addition, we present results for the Born cross
section in the double-pole approximation (DPA) as described in
ref.~\cite{Accomando:2004de} and in the narrow-width approximation
(NWA), where resonant propagators are replaced according to
\begin{equation}
\frac{1}{(p^2-M_V^2)^2 + M_V^2 \Gamma_V^2} \to \frac{\pi}{M_V\Gamma_V}\,\delta(p^2-M_V^2)\,,
\end{equation}
corresponding to the limit $\Gamma_V/M_V \to 0$.  Both
approximations only include doubly-resonant contributions.  Note that
both the $\gamma^*$ and Z-induced amplitudes are included in the full LO
predictions, while the $\gamma^*$-mediated diagrams are absent in the
approximate results. The relative weak corrections, which are identical
to a level of 1~$\permil$ between DPA and NWA, are listed in column~5.

With the cuts employed until now a sizable fraction of events
corresponds to large $\hat{s}$ but small $|\hat{t}|$, and the Sudakov
approximation is not applicable. Imposing, however, a cut on the
rapidity difference between the reconstructed Z bosons, $|\Delta
y_{\PZ\PZ}|<3$, removes events with small scattering angle and decreases
the total rates by roughly $30\%$ at high
$M_{\mathrm{inv}}(4l)$. Moreover, it leads to enhanced EW corrections,
as shown in table~\ref{ta:zz4l_sudakov}. We point out that the results
of table~\ref{ta:zz4l_sudakov} are in good agreement (better than 2\%)
with those presented in table~3 of ref.~\cite{Accomando:2004de}, despite
the fact that we do not include QED corrections and corresponding
non-factorizable contributions which arise in the DPA.
\begin{table}
\begin{equation}
\begin{array}{|c||c|c|c|c|c|}
\hline
 \multicolumn{6}{|c|}{\mathrm{pp} \to \mathrm{(Z/\gamma^*)(Z/\gamma^*)} + X \to
    \Pe^+\Pe^-\mu^+\mu^- + X} \\
\hline
M_{\mathrm{inv}}^{\mathrm{cut}}(4l)/\mathrm{GeV} & \sigma^{\mathrm{naive}}_{\mathrm{LO}}/\mathrm{pb} & \sigma_{\mathrm{LO}}^{\mathrm{CMS}}/\mathrm{pb} &
\sigma_{\mathrm{LO}}^{\mathrm{DPA}}/\mathrm{pb}& \sigma_{\mathrm{LO}}^{\mathrm{NWA}}/\mathrm{pb} &
\delta^{\mathrm{DPA}}_{\mathrm{weak}}/\% 
 \nonumber \\
\hline
\hline 
\multicolumn{6}{|c|}{\mbox{LHC14}} \\
\hline
200&0.835\times10^{-2}&0.835\times10^{-2}&0.815\times10^{-2}&0.875\times10^{-2}&-5.4   \nonumber \\
300&0.239\times10^{-2}&0.239\times10^{-2}&0.233\times10^{-2}&0.249\times10^{-2}&-8.1   \nonumber \\
400&0.987\times10^{-3}&0.987\times10^{-3}&0.966\times10^{-3}&1.035\times10^{-3}&-10.2  \nonumber \\
500&0.484\times10^{-3}&0.484\times10^{-3}&0.473\times10^{-3}&0.508\times10^{-3}&-12.4  \nonumber \\
600&0.261\times10^{-3}&0.261\times10^{-3}&0.256\times10^{-3}&0.275\times10^{-3}&-14.5\nonumber \\
700&0.151\times10^{-3}&0.151\times10^{-3}&0.148\times10^{-3}&0.159\times10^{-3}&-16.6 \nonumber \\
800&0.920\times10^{-4}&0.920\times10^{-4}&0.901\times10^{-4}&0.971\times10^{-4}&-18.8 \nonumber \\
900&0.584\times10^{-4}&0.584\times10^{-4}&0.572\times10^{-4}&0.617\times10^{-4}&-20.8 \nonumber \\
1000&0.384\times10^{-4}&0.384\times10^{-4}&0.376\times10^{-4}&0.406\times10^{-4}&-22.9 \nonumber \\
\hline
 \multicolumn{6}{|c|}{\mbox{LHC8}} \\
 \hline
 200&0.445\times10^{-2}&0.445\times10^{-2}&0.435\times10^{-2}&0.466\times10^{-2}&-5.3   \nonumber \\
 300&0.120\times10^{-3}&0.120\times10^{-3}&0.117\times10^{-2}&0.125\times10^{-3}&-7.7   \nonumber \\
 400&0.463\times10^{-3}&0.463\times10^{-3}&0.452\times10^{-3}&0.484\times10^{-3}&-9.4   \nonumber \\
 500&0.210\times10^{-3}&0.210\times10^{-3}&0.206\times10^{-3}&0.221\times10^{-3}&-11.2  \nonumber \\
 600&0.105\times10^{-3}&0.105\times10^{-3}&0.103\times10^{-3}&0.110\times10^{-3}&-12.9 \nonumber \\
 700&0.557\times10^{-3}&0.557\times10^{-3}&0.544\times10^{-3}&0.586\times10^{-3}&-14.8 \nonumber \\
 800&0.309\times10^{-4}&0.309\times10^{-4}&0.302\times10^{-4}&0.326\times10^{-4}&-16.6 \nonumber \\
 900&0.178\times10^{-4}&0.178\times10^{-4}&0.174\times10^{-4}&0.186\times10^{-4}&-18.5 \nonumber \\
 1000&0.106\times10^{-4}&0.106\times10^{-4}&0.103\times10^{-4}&0.111\times10^{-4}&-20.4 \nonumber \\
 \hline
\end{array}
\end{equation}
\caption{\label{ta:zz4l} LO cross section for Z-boson pair
  production with 4-lepton final states and corresponding weak
  corrections at the LHC for different cut values of the 4-lepton invariant mass.}
\end{table}

\begin{table}
\begin{equation}
\begin{array}{|c||c|c|c|c|c|}
\hline
 \multicolumn{6}{|c|}{\mathrm{pp} \to \mathrm{(Z/\gamma^*)(Z/\gamma^*)} + X \to
   \Pe^+\Pe^-\mu^+\mu^- + X,\;|\Delta y_{\mathrm{ZZ}}|<3} \\
\hline
M_{\mathrm{inv}}^{\mathrm{cut}}(4l)/\mathrm{GeV} & \sigma^{\mathrm{naive}}_{\mathrm{LO}}/\mathrm{pb} & \sigma_{\mathrm{LO}}^{\mathrm{CMS}}/\mathrm{pb} &
\sigma_{\mathrm{LO}}^{\mathrm{DPA}}/\mathrm{pb}& \sigma_{\mathrm{LO}}^{\mathrm{NWA}}/\mathrm{pb} &
\delta^{\mathrm{DPA}}_{\mathrm{weak}}/\%\nonumber \\
\hline
\hline 
\multicolumn{6}{|c|}{\mbox{LHC14}} \\
\hline
200&0.815\times10^{-2}&0.815\times10^{-2}&0.795\times10^{-2}&0.855\times10^{-2}&-5.4   \nonumber \\
300&0.219\times10^{-2}&0.219\times10^{-2}&0.214\times10^{-2}&0.229\times10^{-2}&-8.4   \nonumber \\
400&0.791\times10^{-3}&0.791\times10^{-3}&0.770\times10^{-3}&0.828\times10^{-3}&-11.6  \nonumber \\
500&0.326\times10^{-3}&0.326\times10^{-3}&0.319\times10^{-3}&0.343\times10^{-3}&-15.9  \nonumber \\
600&0.168\times10^{-3}&0.168\times10^{-3}&0.164\times10^{-3}&0.177\times10^{-3}&-19.3 \nonumber \\
700&0.962\times10^{-4}&0.962\times10^{-4}&0.941\times10^{-4}&1.017\times10^{-4}&-22.3 \nonumber \\
800&0.587\times10^{-4}&0.587\times10^{-4}&0.575\times10^{-4}&0.621\times10^{-4}&-24.9 \nonumber \\
900&0.374\times10^{-4}&0.374\times10^{-4}&0.367\times10^{-4}&0.397\times10^{-4}&-27.4 \nonumber \\
1000&0.247\times10^{-4}&0.247\times10^{-4}&0.242\times10^{-4}&0.262\times10^{-4}&-29.7 \nonumber \\
\hline
\end{array}
\end{equation}
\caption{\label{ta:zz4l_sudakov} LO cross section for Z-boson pair
  production with 4-lepton final states and corresponding weak
  corrections at LHC14 for different cut values of the 4-lepton invariant mass in
  the Sudakov regime with $|\Delta y_{\mathrm{ZZ}}|<3$.}
\end{table}

\subsection{Real-radiation contributions}
\label{se:real_rad}
\begin{figure}
\begin{center}
\includegraphics[width = 1.0\textwidth]{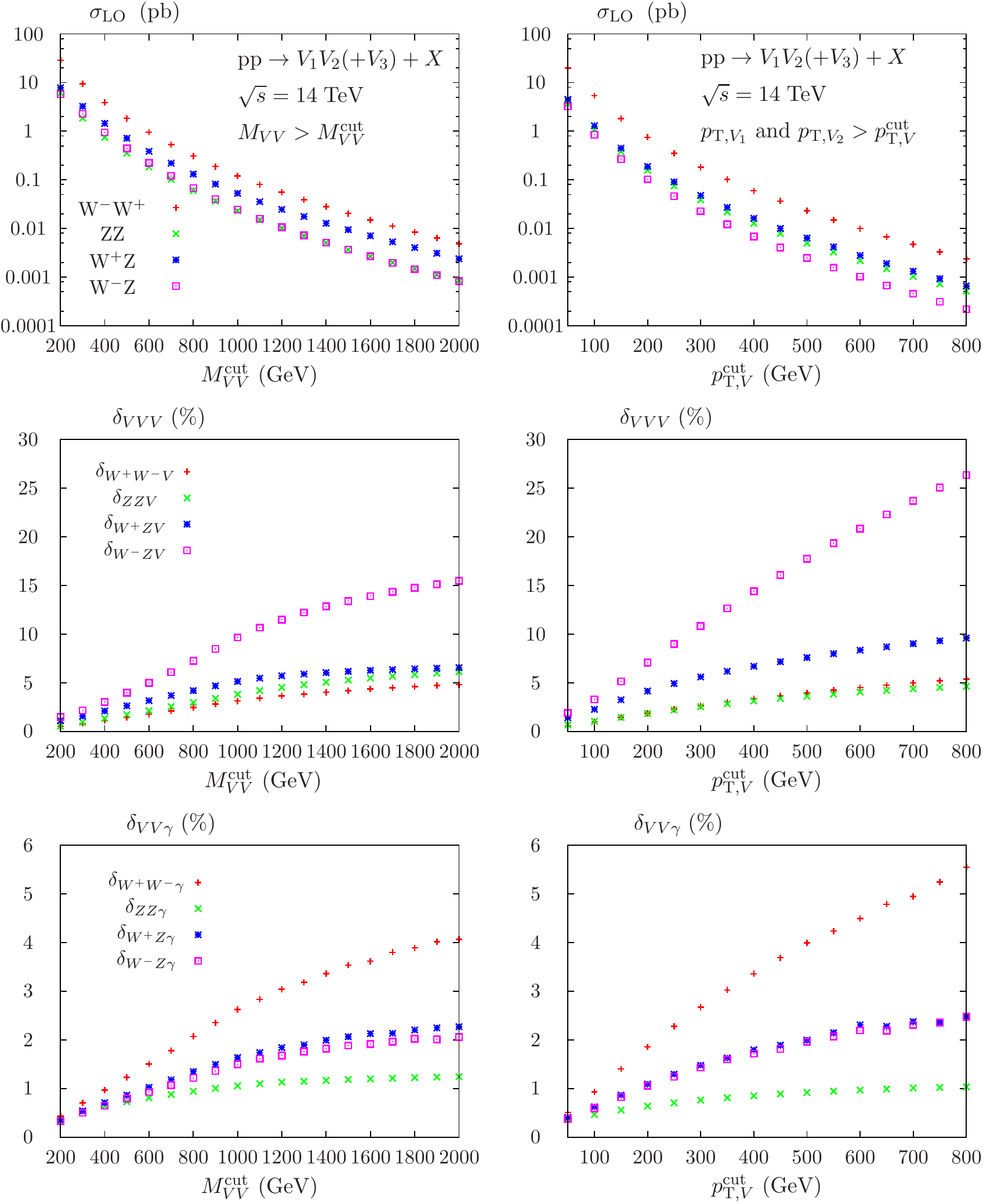}
\end{center}
\caption{\label{fi:real_L14} Integrated LO cross sections (top) and
  relative corrections at the LHC14 due to radiation of one additional
  massive vector boson (center) and hard-photon radiation (bottom)
  evaluated with our default setup for different cuts on the invariant
  mass (left)/transverse momenta (right) of the final-state bosons.}
\end{figure}

\begin{figure}
\begin{center}
\includegraphics[width = 1.0\textwidth]{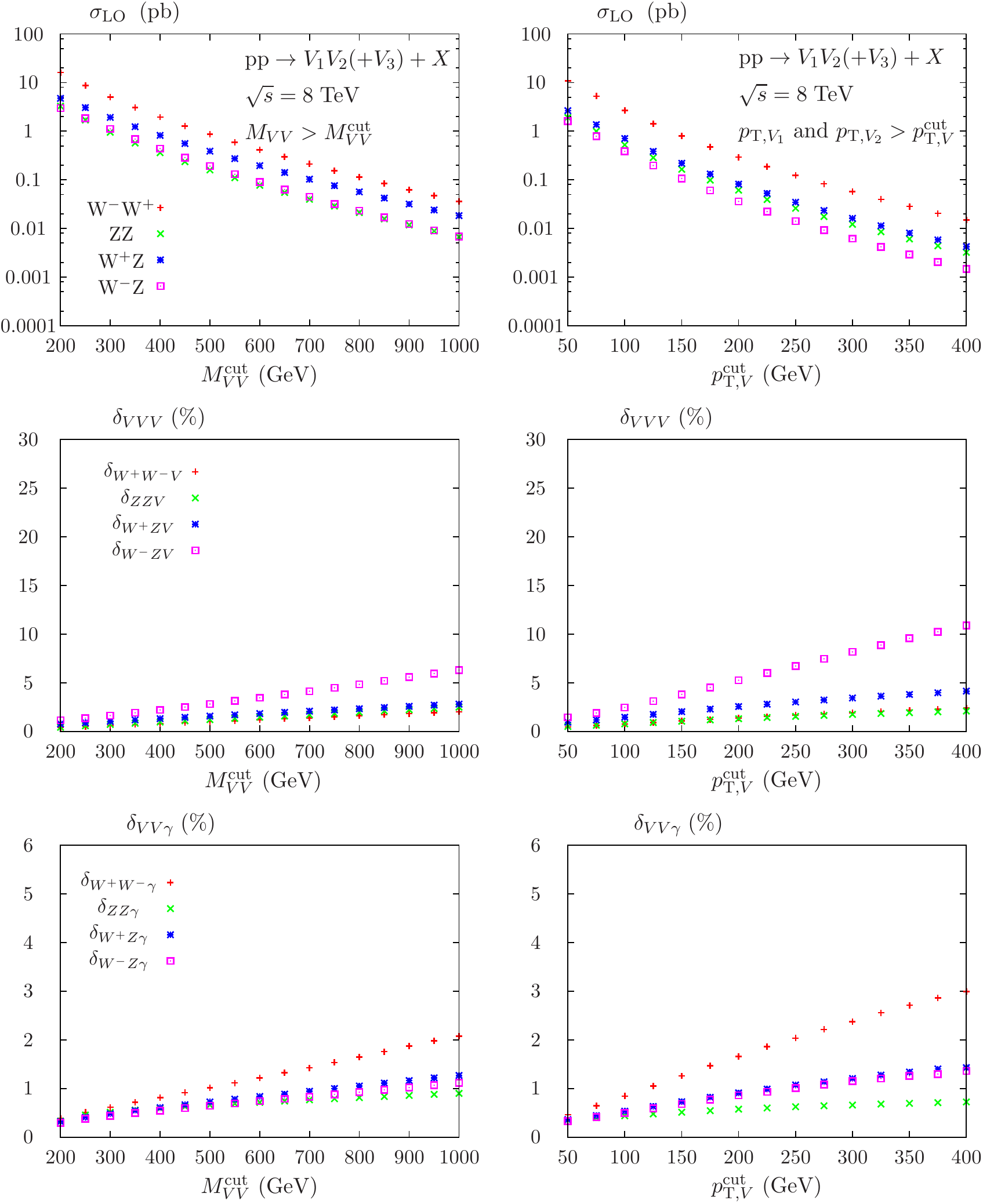}
\end{center}
\caption{\label{fi:real_L8} Integrated LO cross sections (top) and
  relative corrections at the LHC8 due to radiation of one additional
  massive vector boson (center) and hard-photon radiation (bottom)
  evaluated with our default setup for different cuts on the invariant
  mass (left)/transverse momenta (right) of the final-state bosons.}
\end{figure}

Let us finally investigate the phenomenological effect of additional
massive-boson radiation as defined in section~\ref{se:real_rad_def}.  
In figures~\ref{fi:real_L14} and~\ref{fi:real_L8} we show the
corresponding relative corrections for LHC14 and LHC8, respectively. As
far as WW, ZZ and W$^+$Z production at LHC14 is concerned, the
contributions, which are always below 10\%, are found to be of minor
importance for the phenomenological analysis, and the effects are even
smaller at LHC8, as expected. In contrast to this, in the case of W$^-$Z
production we observe remarkably large corrections reaching $+25\%$
($+15\%$) at LHC14 (LHC8). However, we point out that this effect,
though sizable, cannot be attributed to large logarithms arising from
the infrared structure of the corresponding squared matrix elements. In
fact, it can be easily understood recalling that the corresponding
real-radiation contributions as defined in section~\ref{se:real_rad_def}
include W$^-$ZW$^+$ production at LO which, compared to W$^-$Z
production is from the beginning enhanced by a factor of 2 due to one
u-quark PDF factor.  To verify this line of argumentation, we have
checked that for the Tevatron the relative corrections for W$^\pm$Z
production indeed coincide, not exceeding the level of $+5\%$ for
$p_{\rT}^\cut = 300$ GeV. Nevertheless, the numerical effects discussed
here can easily (and definitely should be) taken into account in the
experimental analysis of the background.

Turning to the effects of hard photon radiation displayed in the lower
two plots of figures~\ref{fi:real_L14} and~\ref{fi:real_L8},
respectively, we find that the corresponding contributions are moderate
and by far largest for W-pair production, exceeding $+5\%$, while they
are completely irrelevant for Z-pair production.

\section{Conclusions}
We have computed the full one-loop electroweak corrections to on-shell
ZZ, W$^\pm$Z, ZZ and $\gamma\gamma$ production at hadron colliders, for
the first time consistently taking into account all mass
effects. Furthermore, the results presented are not limited to a
particular kinematic regime, allowing for flexible predictions valid in
all regions of phase space. In case of ZZ production we have also
included the leptonic decays and the corresponding weak corrections,
finding good agreement with former computations restricted to Sudakov
kinematics.
The relative corrections are negative and grow with increasing
center-of-mass energy. They are largest for ZZ production, reaching
$-50\%$ at energies accessible at LHC14, and smallest for $\gamma\gamma$
production. As an interesting new finding we observe that the relative
corrections are not only sizable in the Sudakov regime as has been shown
before, but may also---in the ZZ case---give significant contributions
at rather low transverse momenta. We have also investigated the effect
of massive boson radiation processes which may be considered as
background to vector-boson pair production depending on the details of
the experimental setup. The effects are smaller than expected from naive
partonic considerations and may easily be included in the experimental
studies. Our predictions rely on the experimental reconstruction of the
intermediate bosons. In the future leptonic decays of the vector bosons
will be included also in the analysis of WW and WZ production
together with the corresponding $\O(\alpha)$ corrections to allow for a
more realistic event definition.

\subsection*{Acknowledgements}
This work has been supported by ``Strukturiertes Promotionskolleg
Elementarteilchen- und Astroteilchenphysik'', SFB TR9 ``Computational
and Particle Physics'' and BMBF Contract 05HT4VKATI3.

\end{document}